\documentclass[prl,nofootinbib,showpacs]{revtex4}

\usepackage{epsf}

\begin{document}

%\draft

%\tighten

\title
{\bf Superconductivity in the ``Hot Spots'' Model of the Pseudogap
State: Impurity Scattering and Phase Diagram }
\author{N.A.Kuleeva\footnote{E-mail:\ strigina@iep.uran.ru},\ 
E.Z.Kuchinskii\footnote{kuchinsk@iep.uran.ru},\ 
M.V.Sadovskii\footnote{E-mail:\ sadovski@iep.uran.ru}}

%\address
\affiliation
{Institute for Electrophysics,\ Russian Academy of Sciences,\  Ekaterinburg,\ 
620016, Russia}

%\date{} 

%\maketitle

%\begin{center}
%{\sl Submitted to JETP  }\\
%{\sl cond-mat/04}
%\end{center}

\begin{abstract}

We analyze the anomalies of superconducting state (both $s$ and ¨ $d$-wave
pairing) in the model of the pseudogap state induced by Heisenberg spin 
fluctuations of antiferromagnetic short -- range order, and based on the
scenario of strong scattering near the ``hot spots'' on the Fermi surface.
We present microscopic derivation of Ginzburg -- Landau expansion, taking
into account {\em all} Feynman graphs of perturbation theory for electron
interaction with fluctuations of short -- range order and in the ``ladder''
approximation for electron scattering by normal (nonmagnetic) impurities.
We determine the dependence of superconducting critical temperature $T_c$
and other characteristics of a superconductor on the parameters of the
pseudogap and impurity scattering. It is shown that within this model it is
possible to explain the typical phase diagram of high -- temperature
superconductors.

\end{abstract}


\maketitle

%\newpage

%\narrowtext

\section{Introduction}

One of the main problems in the physics of high --  temperature 
superconducting cuprates remains the theoretical understanding of typical
phase diagram of these compounds \cite{Lor}. Especially interesting is the
clarification of the nature of the pseudogap state which is observed in a wide
region of temperatures and concentration of carriers \cite{MS}, and is
obviously crucial for the formation of electronic properties both in normal
and superconducting states. Despite continuing discussions on the nature of
the pseudogap, it seems that the preferable ``scenario'' of its formation is,
in our opinion, based on the on the model of strong scattering of current 
carriers by antiferromagnetic (AFM, SDW) spin fluctuations of short -- range 
order\footnote{Similar charge (CDW) fluctuations also cannot be excluded.}
\cite{MS,Pines}. In momentum representation this scattering transfers 
momenta of the order of ${\bf Q}=(\frac{\pi}{a},\frac{\pi}{a})$ 
($a$ -- lattice constant of two -- dimensional lattice) 
and leads to the ``precursors'' of spectrum transformation, taking place
after the establishment of long -- range AFM order (period doubling).
As a result we obtain non -- Fermi liquid like behavior (dielectrization)
of spectral density in the vicinity of the so called ``hot spots'' on the
Fermi surface, appearing at intersections of this surface with ``would be''
antiferromagnetic Brillouin zone boundary \cite{MS}.

Within this approach a simplified model of the pseudogap state was actively
studied \cite{Sch,KS}, under the assumption that the scattering by dynamic
spin fluctuations can be reduced (which is valid for high enough temperatures)
to a static Gaussian random field (quenched disorder) of pseudogap fluctuations
with characteristic scattering vector from the vicinity of ${\bf Q}$, the
width of which is determined by the inverse correlation length of short --
range order $\kappa=\xi^{-1}$. The review of this approach with applications to
the properties of the normal state and for some oversimplified models of
pseudogap fluctuations influence on superconductivity can be found in 
Ref. \cite{MS}.

In a recent paper \cite{KSS} we have presented microscopic derivation of
Ginzburg -- Landau expansion\footnote{Similar analysis based on Gorkov's
equations was given in Ref. \cite{KK}} and studied the influence of pseudogap
fluctuations in the model of ``hot spots'' on the main characteristics of
superconducting state (both for $s$-wave and $d$-wave pairing), forming ``on the
background'' of these fluctuations. In this paper we have considered slightly
oversimplified version of our model, where Heisenberg -- like spin fluctuations
were replaced either by Ising -- like, or by spin -- independent charge (CDW)
fluctuations. It was shown that such fluctuations of ``dielectric'' nature
generally suppress superconductivity, leading to the drop of superconducting
transition temperature and of specific heat discontinuity at the transition,
as well as to a number of other anomalies. We also discovered two possible
classes of superconducting order parameter interaction with pseudogap
fluctuations, leading to significantly different scales of suppression of
superconducting state.

The aim of the present paper is the generalization of the approach of 
Ref. \cite{KSS} to the ``realistic'' case of Heisenberg spin fluctuations,
as well as the account of (nonmagnetic) impurity (disorder) scattering
influence on superconductivity in the pseudogap state. It will be shown
that this model allows for semiquantitative modeling of the typical phase
diagram of a high -- temperature superconductor.

\section{``Hot Spots'' Model and Recursion Procedure for Green's function 
and Vertices.}

Main results of our approach to the ``hot spots'' model and calculation of a
single -- particle Green's function in this model were described in detail
in Refs. \cite{Sch,KS}, methods of calculation for the appropriate vertex
parts were discussed in Refs. \cite{KSS,SS}. Here we only present the main
equations and definitions, giving a short description of changes necessary
to take into account the spin structure of interaction in Heisenberg model
of antiferromagnetic fluctuations.

Effective interaction of electrons with spin fluctuations in the model of
``nearly antiferromagnetic'' Fermi -- liquid \cite{Sch} is described by dynamic
susceptibility, characterized by experimentally determined correlation length
$\xi$ and frequency $\omega_{sf}$ of spin fluctuations, which can depend both
on concentration of carriers (and in case of $\xi$ also on temperature), 
Both dynamic susceptibility and effective interaction are peaked (in momentum
representation) in the vicinity of ${\bf Q}=(\pi/a,\pi/a)$, which leads to the
appearance of two types of quasiparticles -- ``hot'' one, with momenta close
to the points on the Fermi surface, connected by scattering vectors 
$\sim{\bf Q}$, and ``cold'' one, with the momenta close to the parts of the
Fermi surface, surrounding diagonals of the Brillouin zone \cite{MS,Sch,KS}.

\begin{figure}
\epsfxsize=16cm
\epsfysize=20cm
\epsfbox{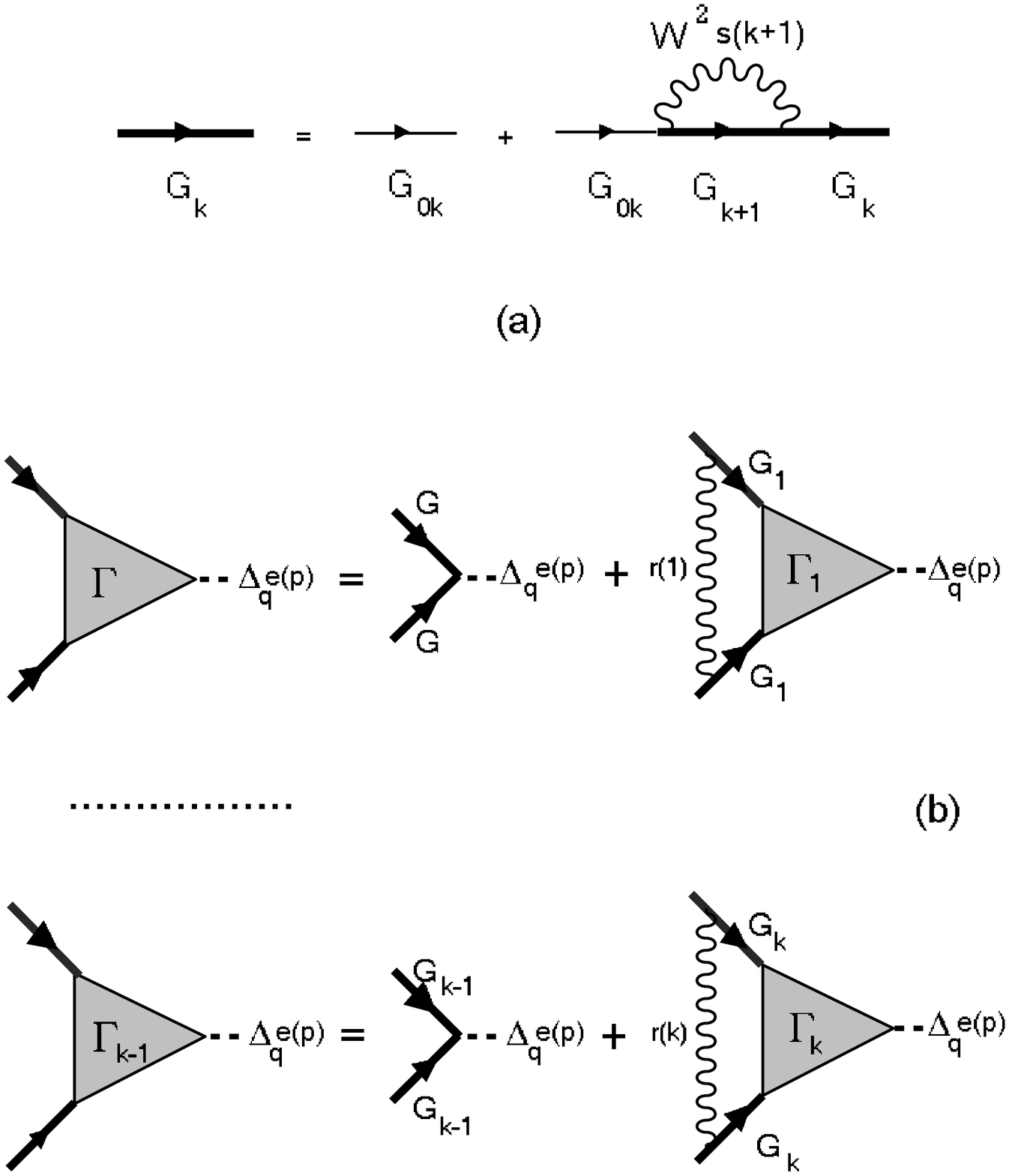}
\caption{Recurrence equations for Green's function (a) and ``triangular''
vertex (b).}
\label{Dyson}
\end{figure}

For high enough temperatures $2\pi T\gg \omega_{sf}$ and we can neglect spin
dynamics \cite{Sch}. Electron interaction with spin (pseudogap) fluctuations
reduces then to elastic scattering by appropriate static Gaussian random field.
In this model it is possible to introduce a simplified form of effective
interaction (correlator of this random field) \cite{Sch,KS}, which allows to
perform full summation of the Feynman perturbation series, and obtain the
following recursion procedure determining the single -- electron Green's
function:
\begin{equation}
G_k(\varepsilon_n{\bf p})=\frac{1}{i\varepsilon_n-\xi_{k}({\bf p})+
ikv_k\kappa-\Sigma_{k}(\varepsilon_n{\bf p})}
\label{Gk}
\end{equation}
\begin{equation}
\Sigma_k(\varepsilon_n{\bf p})=W^2s(k+1)G_{k+1}(\varepsilon_n{\bf p})
\label{Sig_k}
\end{equation}
which is depicted as a symbolic ``Dyson like'' equation in 
Fig. \ref{Dyson} (a), where we have also introduced:
\begin{equation}
G_{0k}(\varepsilon_n{\bf p})=\frac{1}{i\varepsilon_n-\xi_k({\bf p})+
ikv_k\kappa}
\label{G0k}
\end{equation}
Here $\kappa=\xi^{-1}$ is the inverse correlation length of pseudogap
fluctuations, $\varepsilon_n=2\pi T(n+1/2)$ (for definiteness we assume here 
$\varepsilon_n>0$), 
\begin{equation} \xi_k({\bf p}) 
=\left\{\begin{array}{ll} \xi_{{\bf p}+{\bf Q}} & \mbox{for odd $k$} \\ 
\xi_{\bf p} & \mbox{for even $k$}
\end{array} \right.
\label{xik}
\end{equation}
\begin{equation}
v_k=\left\{\begin{array}{ll}
|v_x({\bf p}+{\bf Q})|+|v_y({\bf p}+{\bf Q})| & \mbox{for odd $k$} \\
|v_x({\bf p})|+|v_y({\bf p})| & \mbox{for even $k$}
\end{array} \right.
\label{Vk}
\end{equation}
where ${\bf v}({\bf p})=\frac{\partial\xi_{\bf p}}{\partial {\bf p}}$ is the
velocity of a free quasiparticle with the spectrum $\xi_{\bf p}$, which is 
taken in a standard form \cite{Sch}:
\begin{equation}
\xi_{\bf p}=-2t(\cos p_xa+\cos p_ya)-4t^{'}\cos p_xa\cos p_ya - \mu
\label{spectr}
\end{equation}
where $t$ is the transfer integral between nearest neighbors, 
and $t'$ --- between second nearest neighbors on the square lattice, 
$a$ is the lattice constant, $\mu$ is the chemical potential.

Parameter $W$ with dimension of energy defines the effective width of the
pseudogap. In the model of Heisenberg spin fluctuations it can be written as
\cite{Sch}:
\begin{equation}
W^2=g^2\frac{<{\bf S}_i^2>}{3}=g^2<(n_{i\uparrow}-n_{i\downarrow})^2>
\label{W2}
\end{equation}
where $g$ is an interaction constant of electrons with spin fluctuations,
$<{\bf S}_i^2>$ is the average square of the spin on a lattice site,
$n_{i\uparrow}$,\ $n_{i\downarrow}$ -- operators of electron number at a site,
with given spin directions. It is clear that similarly to correlation length
$\xi$, the value of $W$, in our semiphenomenological approach
\cite{Sch,KS}, is also some function of carrier concentration (and temperature),
to be determined from the experiment.

The value of $s(k)$ is determined by combinatorics of Feynman diagrams.
For the simplest case of commensurate charged (CDW) pseudogap fluctuations :
\begin{equation}
s(k)=k
\label{vcomm}
\end{equation}
while for the most interesting case of Heisenberg spin (SDW) fluctuations
\cite{Sch}\footnote{Detailed analysis of diagram combinatorics for Heisenberg
model of spin fluctuations is given in the Appendix.}:
\begin{equation}
s(k)=\left\{\begin{array}{cc}
\frac{k+2}{3} & \mbox{for odd $k$} \\
\frac{k}{3} & \mbox{for even $k$}
\end{array} \right.
\label{sHeis}
\end{equation}
Conditions of applicability of our approximations were discussed in detail
in Refs. \cite{Sch,KS}. 

Remarkable property of our model is the possibility of complete summation of
all Feynman diagrams for the vertex parts\footnote{Including all diagrams with
intersecting interaction lines.}, describing system response to an arbitrary
external perturbation \cite{SS}. Here we just give appropriate recursion
relations for ``triangular'' vertex in Cooper channel, similar to those derived
in Ref. \cite{KSS} in particle -- particle channel, and describing the response
to an arbitrary fluctuation of superconducting order parameter (gap):
\begin{equation}
\Delta({\bf p},{\bf q})=\Delta_{\bf q}e({\bf p})
\label{Deltq}
\end{equation}
where the symmetry factor, corresponding to the type (symmetry) of the
pairing, takes the following form:
\begin{equation}
e({\bf p})=
\left\{
\begin{array}{ll}
1 & (\mbox{ $s$-wave pairing})\\ 
\cos p_xa-\cos p_ya & (\mbox{ $d_{x^2-y^2}$-wave pairing})
\end{array}.
\right.
\label{ephi}
\end{equation}
and we always assume singlet pairing. The vertex of interest to us can be 
written as:
\begin{equation}
\Gamma(\varepsilon_n,-\varepsilon_n,{\bf p},{\bf -p+q})\equiv
\Gamma_{\bf p}(\varepsilon_n,-\varepsilon_n,{\bf q})e({\bf p})
\label{Gamephi}
\end{equation}
Then $\Gamma_{\bf p}(\varepsilon_n,-\varepsilon_n,{\bf q})$ is determined from
recursion procedure of the following form:
\begin{eqnarray}
\Gamma_{{\bf p}k-1}(\varepsilon_n,-\varepsilon_n,{\bf q})=1 \pm 
W^2r(k)G_k(\varepsilon_n,{\bf p+q}) G_k(-\varepsilon_n,{\bf p})
\Biggl\{1+\nonumber\\
+\frac{2ik\kappa v_k}
{G^{-1}_{k}(\varepsilon_n,{\bf p+q})-G^{-1}_{k}(-\varepsilon_n,{\bf p})
-2ik\kappa v_k}\Biggr\}\Gamma_{{\bf p}k}(\varepsilon_n,-\varepsilon_n,{\bf q}) 
\label{Gamma} 
\end{eqnarray}
which is shown diagrammatically in Fig. \ref{Dyson} (b). ``Physical'' vertex is
given by $\Gamma_{{\bf p}k=0}(\varepsilon_n,-\varepsilon_n,{\bf q})$. 
Additional combinatorial factor $r(k)=s(k)$ for the simplest case of charged
(or Ising -- like spin) pseudogap fluctuations, analyzed in Ref. \cite{KSS}. 
For the most interesting case of Heisenberg spin (SDW) fluctuations, analyzed 
below, this factor is givwn by \cite{Sch} (Cf. also Appendix):
\begin{equation}
r(k)=\left\{\begin{array}{ll}
k & \mbox{for even $k$} \\
\frac{k+2}{9} & \mbox{for odd $k$}
\end{array} \right.
\label{rk}
\end{equation}
The choice of the sign before $W^2$ in the r.h.s of Eq. (\ref{Gamma}) depends
on the symmetry of superconducting order parameter and the type of pseudogap
fluctuations \cite{KSS} (see details in the Appendix). Different variants are
given in Table I.
\begin{center}
{\sl Table I.\ The choice of the sign in the recursion procedure for the
vertex part.}
\begin{tabular}{|c|c|c|c|}
\hline
Pairing & CDW - fluctuations & SDW - fluctuations (Ising) 
&  SDW - fluctuations (Heisenberg) \\ 
\hline 
$s$ & $+$ & $-$ & $+$     \\ 
\hline 
$d$ & $-$ & $+$ & $-$  \\ 
\hline
\end{tabular}
\end{center}
It particular, from this Table we can see, that in the most interesting 
case of $d$-wave pairing and Heisenberg spin fluctuations, we must take this 
sign ``$-$'', so that alternating signs appear in our recursion procedure for
the vertex part. However, for the case of $s$-wave pairing and the same type
of pseudogap fluctuations, we must take ``$+$'', and there is no sign
alternation in the recursion procedure. It Ref. \cite{KSS} it was shown
(for other cases shown in Table I), that this difference in the signs behavior 
of the recursion procedure for the vertex part leads to two cases of 
qualitatively different behavior of all the main characteristics of a 
superconductor.

\section{Impurity Scattering.}

\begin{figure}
\epsfxsize=16cm
\epsfysize=20cm
\epsfbox{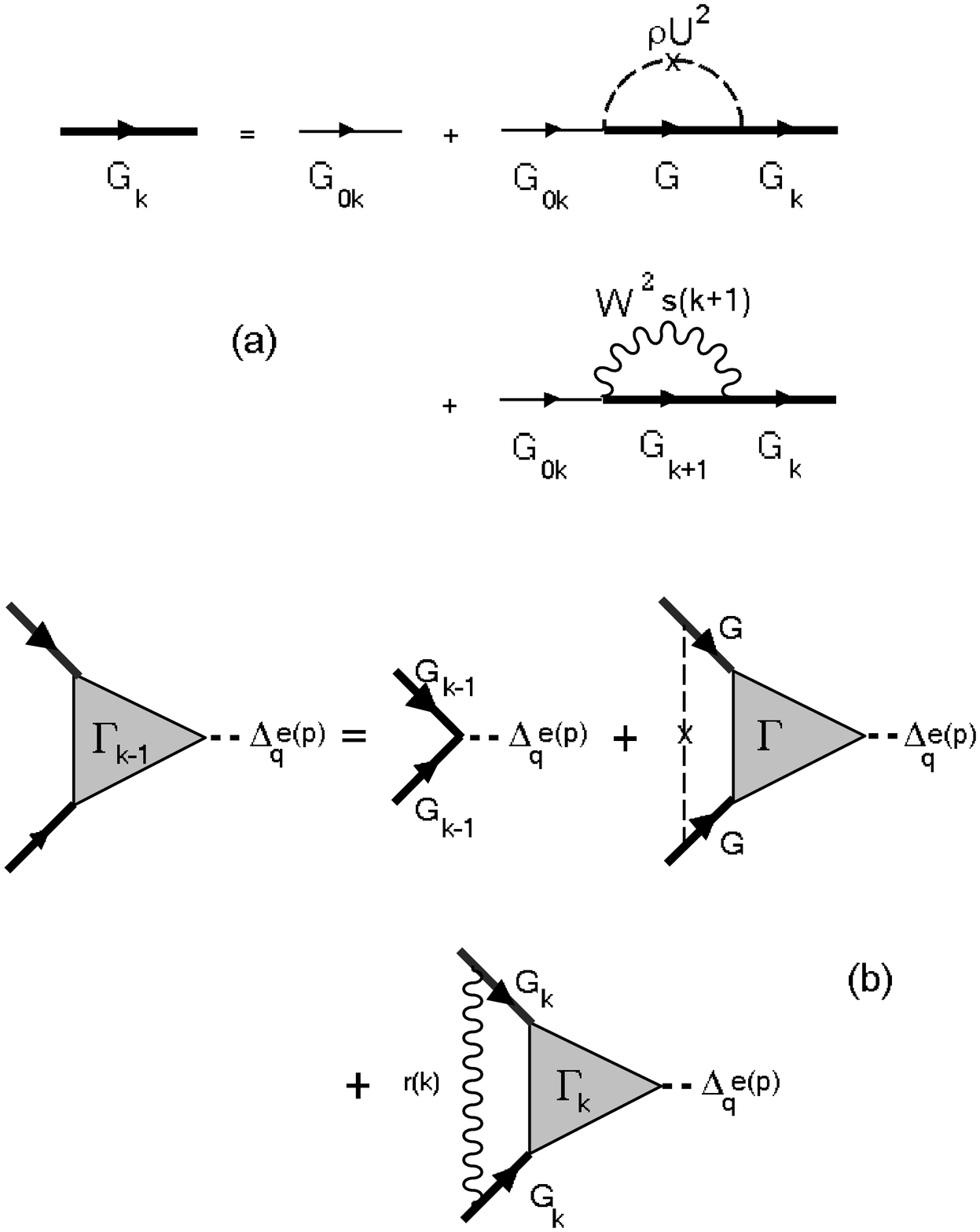}
\caption{Recurrence equations for Green's function (a) and ``triangular'' 
vertex (b) including scattering by random impurities.}
\label{Dysonimp}
\end{figure}

Scattering of electrons by normal (nonmagnetic) impurities can easily be taken
into account in self -- consistent Born approximation, writing down 
``Dyson'' equation for the single -- electron Green's function, shown
graphically in Fig. \ref{Dysonimp} (a), where in addition to contribution
shown in Fig. \ref{Dyson} (a), we have included the standard self -- energy
part due to impurity scattering \cite{AGD}. As a result we obtain the 
recursion relation for the Green's function as:
\begin{equation}
G_k(\varepsilon_n{\bf p})=\frac{1}{G^{-1}_{0k}(\varepsilon_n{\bf p})-
\rho U^2\sum_{\bf p}G(\varepsilon_n{\bf p})-W^2s(k+1)G_{k+1}(\varepsilon_n
{\bf p})}
\label{Dysrecim}
\end{equation}
where $\rho$ is concentration of impurities with point -- like potential
$U$, and impurity contribution to self -- energy contains fully dressed
Green's function $G(\varepsilon_n{\bf p})=G_{k=0}(\varepsilon_n{\bf p})$, 
to be determined self -- consistently by our recursion procedure.
As usual \cite{AGD}, contribution to this self -- energy from real part of the 
Green's function is reduced to insignificant renormalization of the chemical
potential, so that (\ref{Dysrecim}) can be written as:
\begin{equation}
G_k(\varepsilon_n{\bf p})=\frac{1}{i(\varepsilon_n -
\rho U^2\sum_{\bf p}ImG(\varepsilon_n{\bf p})+kv_k\kappa)-\xi_k({\bf p})
-W^2s(k+1)G_{k+1}(\varepsilon_n{\bf p})}
\label{Dysrecimp}
\end{equation}
Thus, in comparison with impurity free case, we obtain, in fact, a simple
substitution (renormalization):
\begin{eqnarray}
\varepsilon_n\to\varepsilon_n-\rho U^2\sum_{\bf p}ImG(\varepsilon_n{\bf p})
\equiv\varepsilon_n\eta_{\epsilon}
\label{reneps}
\\
\eta_{\epsilon}=1-\frac{\rho U^2 }{\varepsilon_n} \sum_{\bf p}ImG(\varepsilon_n{\bf p})
\label{etaeps}
\end{eqnarray}
If we do not perform fully self -- consistent calculation of impurity
self -- energy, in the simplest approximation we just have:
\begin{eqnarray}
\varepsilon_n\to\varepsilon_n-\rho U^2\sum_{\bf p}ImG_{00}
(\varepsilon_n{\bf p}) \equiv\varepsilon_n\eta_{\epsilon} 
=\varepsilon_n+\gamma_0 sign\varepsilon_n
\label{renepsi} 
\\ 
\eta_{\epsilon}=1+\frac{\gamma_0}{|\varepsilon_n|}
\label{etaepsi}
\end{eqnarray}
where $\gamma_0=\pi\rho U^2N_0(0)$ is the standard Born impurity scattering
rate \cite{AGD} ($N_0(0)$ --- density of states of free electrons on the
Fermi level).

For ``triangular'' vertices of interest to us, the recurrence relation with the
account of impurity scattering is shown graphically in Fig. \ref{Dysonimp} (b). 
For the vertex, describing interaction with the fluctuation of superconducting
order parameter (\ref{Deltq}) with $d$-wave symmetry (\ref{ephi}) this equation
is significantly simplified, because the contribution of second diagram in 
the r.h.s. of Fig. \ref{Dysonimp} (b) is actually zero, because of validity of
$\sum_{\bf p}e({\bf p})=0$ (cf. discussion of similar situation in Ref.
\cite{PS}). Then the recurrence equation for the vertex has the form of
Eq. (\ref{Gamma}), where $G_k(\pm\varepsilon_n{\bf p})$ should be taken from
Eqs. (\ref{Dysrecim}),\ (\ref{Dysrecimp}), i.e. just replaced by dressed
by impurity scattering, as defined in Fig. \ref{Dysonimp} (a). For the vertex
describing interaction with fluctuations of the order parameter with
$s$-wave symmetry, we have the following equation: 
\begin{eqnarray} 
&&\Gamma_{{\bf p}k-1}(\varepsilon_n,-\varepsilon_n,{\bf q})=1 + 
\rho U^2\sum_{\bf p}G(\varepsilon_n,{\bf p+q})G(-\varepsilon_n,{\bf p})
\Gamma_{\bf p}(\varepsilon,-\varepsilon_n,{\bf q})\pm 
\label{Gammaimp} \\
&&\pm W^2r(k)G_k(\varepsilon_n,{\bf p+q}) G_k(-\varepsilon_n,{\bf p})
\Biggl\{1+\frac{2ik\kappa v_k}
{G^{-1}_{k}(\varepsilon_n,{\bf p+q})-G^{-1}_{k}(-\varepsilon_n,{\bf p})
-2ik\kappa v_k}\Biggr\}\Gamma_{{\bf p}k}(\varepsilon_n,-\varepsilon_n,{\bf q}) 
\nonumber
\end{eqnarray}
where for $G_k(\pm\varepsilon_n{\bf p})$ we again use Eqs. (\ref{Dysrecim}),\ 
(\ref{Dysrecimp}), and the sign before  $W^2$ is determined in accordance with
the rules formulated above. The difference with the case of the vertex for
$d$-wave symmetry is the appearance of second term in the r.h.s. of
Eq. (\ref{Gammaimp}), which reduces to the substitution:
\begin{equation}
1\to\eta_{\Gamma}=1+\rho U^2\sum_{\bf p}G(\varepsilon_n,{\bf p+q})
G(-\varepsilon_n,{\bf p})\Gamma_{\bf p}(\varepsilon,-\varepsilon_n,{\bf q})
\label{etagam}
\end{equation}
Thus our self-- consistency procedure can be formulated in this case as 
follows: start from the ``zeroth order'' approximation
$G=G_{00}$,\ $\Gamma_{\bf p}=1$,\ i.e. simply take in Eqs.(\ref{Dysrecimp}),\ 
(\ref{Gammaimp}) $\eta_{\varepsilon}=\eta_{\Gamma}=1-\rho U^2/\varepsilon_n 
\sum_{\bf p}ImG_{00}(\varepsilon_n{\bf p})$. Perform all necessary recursions
(starting from some big enough value of $k$) and determine new values for
$G=G_{k=0}$ and $\Gamma_{\bf p}=\Gamma_{k=0}$. Calculate again
$\eta_{\varepsilon}$ and $\eta_{\Gamma}$ using (\ref{etaeps}) and 
(\ref{etagam}), use these values in Eqs. (\ref{Dysrecimp}),\ (\ref{Gammaimp}) 
etc.,\ until convergence.

Analyzing the vertex with $d$-wave symmetry we just take $\eta_{\Gamma}=1$ 
at all stages of calculation. In fact, in this case there is no need for
achieving full self -- consistency over impurity scattering, as it leads only
to some small corrections to the results of non self -- consistent
calculation using the simplest substitution (\ref{renepsi}) \cite{KK}.

\section{Superconducting Transition Temperature and Gizburg -- Landau
Coefficients.}

\begin{figure}
\epsfxsize=12cm
\epsfbox{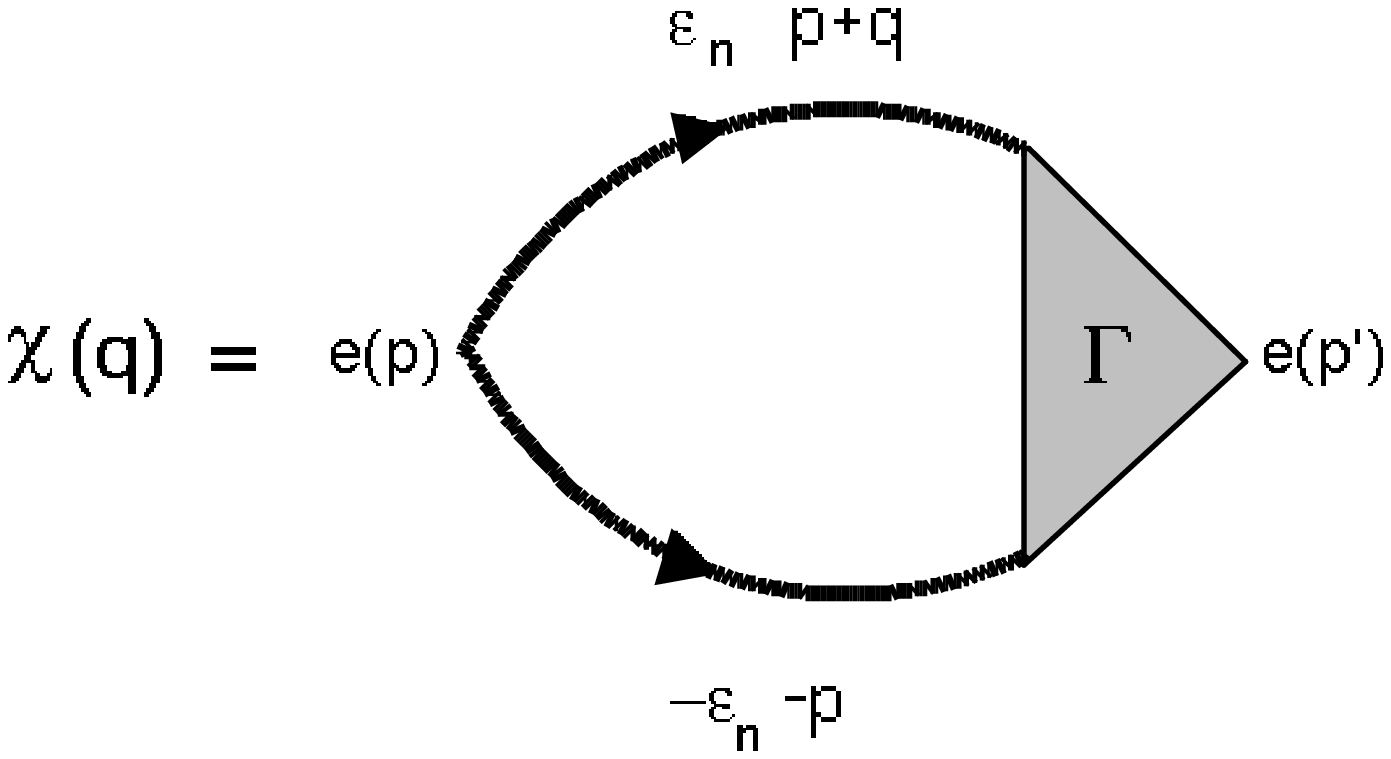}
\caption{Diagrammatic representation of the generalized Cooper channel
susceptibility $\chi ({\bf q})$.}
\label{loop}
\end{figure}

Critical temperature of superconducting transition is determined from the
equation for Cooper instability of the normal phase:
\begin{equation}
1-V\chi(0;T)=0
\label{coopinst}
\end{equation}
where the generalized Cooper susceptibility is defined by diagram shown in
Fig. \ref{loop} and is given by:
\begin{equation}
\chi({\bf q};T)=-T\sum_{\varepsilon_n}\sum_{\bf p}G(\varepsilon_n{\bf p+q})
G(-\varepsilon_n,-{\bf p})e^2({\bf p})
\Gamma_{\bf p}(\varepsilon_n,-\varepsilon_n,{\bf q})
\label{chiq}
\end{equation}
Pairing constant $V$ is assumed to be non zero in a layer of the width of
$2\omega_c$ around the Fermi level, and determines the ``bare'' transition
temperature $T_{c0}$ in the absence of pseudogap fluctuations from the
standard BCS equation\footnote{We do not discuss the physical nature of
this interaction --- it may be due to exchange by spin fluctuations, phonons,
or by some combination of electron phonon and spin fluctuation mechanism.}:  
\begin{equation} 
1=\frac{2VT}{\pi^2}\sum_{n=0}^{\bar m}\int_{0}^{\pi/a}dp_x\int_{0}^{\pi/a}dp_y 
\frac{e^2({\bf p})}{\xi^2_{\bf p}+\varepsilon^2_n} 
\label{TcBCS} 
\end{equation} 
where $\bar m=[\frac{\omega_c}{2\pi T_{c0}}]$ is dimensionless cutoff in the
sum over Matsubara frequencies. Similar to Ref. \cite{KSS} all calculations
were performed for a typical spectrum of bare quasiparticles (\ref{spectr}) 
for different relations between $t$,\ $t'$ and $\mu$. E.g. taking rather
arbitrarily $\omega_c=0.4t$ and $T_{c0}=0.01t$ it is rather easy to find
the value of pairing constant $V$ in (\ref{TcBCS}), responsible for this given
value of $T_{c0}$ for different types of pairing. For example, in the case
of $s$-wave pairing we obtain $\frac{V}{ta^2}=1$, while for the case of
$d_{x^2-y^2}$-wave pairing we get $\frac{V}{ta^2}=0.55$.

To calculate $T_c$ we need only to know the Cooper susceptibility at
$q=0$, and this simplifies calculations considerably \cite{KSS}. In general
case, e.g. to calculate Ginzburg -- Landau coefficients we need
$\chi(q;T)$ at arbitrary (small) $q$.

Ginzburg -- Landau expansion for the difference of free -- energies of
superconducting and normal states, written in a standard form:
\begin{equation}
F_{s}-F_{n}=A|\Delta_{\bf q}|^2
+q^2 C|\Delta_{\bf q}|^2+\frac{B}{2}|\Delta_{\bf q}|^4
\label{GiLa}
\end{equation}
is determined from the usual loop expansion for the free -- energy of electrons,
moving in a random field of order parameter fluctuations (\ref{Deltq}).

It is convenient to normalize GL -- coefficients $A,\ B,\ C$ by their values
in the absence of pseudogap fluctuations \cite{KSS}: 
\begin{equation}
A=A_0K_A;\qquad   C=C_0K_C;\qquad    B=B_0K_B,
\label{ACD}
\end{equation}
where
\begin{eqnarray} 
&&A_0=N_0(0)\frac{T-T_{c}}{T_{c}}<e^2({\bf p})>;\quad 
C_0=N_0(0)\frac{7\zeta(3)}{32\pi^{2}T_c^2}<|{\bf v}({\bf p})|^2e^2({\bf p)}>; 
\nonumber\\ 
&&B_0=N_0(0)\frac{7\zeta(3)}{8\pi^{2}T_c^2}<e^4({\bf p})>,
\label{ACDf}
\end{eqnarray}
and angular brackets denote usual averaging over the Fermi surface:
$<\ldots>=\frac{1}{N_0(0)}\sum_p\delta (\xi_{\bf p})\ldots$,\ 
$N_0(0)$ is again the density of state on the Fermi level for free electrons.

Then we have the following general expressions \cite{KSS}:
\begin{equation} 
K_A=\frac{\chi(0;T)-\chi(0;T_c)}{A_0}
\label{Ka}
\end{equation}
\begin{equation}
K_C=\lim_{q\to 0}\frac{\chi({\bf q};T_c)-\chi(0;T_c)}{q^2C_0}
\label{Kc}
\end{equation}
\begin{equation}
K_B=\frac{T_c}{B_0}\sum_{\varepsilon_n}\sum_{\bf p}e^4({\bf p})
(G(\varepsilon_n{\bf p})G(-\varepsilon_n,-{\bf p}))^2
(\Gamma_{\bf p}(\varepsilon_n,-\varepsilon_n,0))^4
\label{Kb}
\end{equation}
which were used for direct numerical computations.

In the presence of impurities all Green's functions and vertices entering
these expressions are to be calculated using Eqs. (\ref{Dysrecimp}) 
and (\ref{Gammaimp}).

The knowledge of GL -- coefficients allows to determine all major 
characteristics  of a superconductor close to the transition temperature
$T_c$. E.g. the coherence length is defined as:
\begin{equation}
\frac{\xi^2(T)}{\xi_{BCS}^2(T)}=\frac{K_C}{K_A},
\label{xiii}
\end{equation}
where $\xi_{BCS}(T)$ is the value of coherence length in the absence of
pseudogap.

Penetration depth is given by:
\begin{equation}
\frac{\lambda(T)}{\lambda_{BCS}(T)}=
\left(\frac{K_{B}}{K_{A}K_{C}}\right)^{1/2},
\label{lm}
\end{equation}
where we again normalized on the value of $\lambda_{BCS}(T)$ in the absence
of pseudogap fluctuations.

The slope of the upper critical magnetic field close to $T_{c}$, normalized
in a similar way, is given by:  
\begin{equation} 
\frac{\left|\frac{dH_{c2}}{dT}\right|_{T_c}}
{\left|\frac{dH_{c2}}{dT}\right|_{T_{c0}}}= 
\frac{T_{c}}{T_{c0}}\frac{K_A}{K_C}. 
\label{dHc2}
\end{equation}
Relative discontinuity of specific heat at the transition is defined as:
\begin{equation} 
\Delta C=\frac{(C_s-C_n)_{T_c}}{(C_s-C_n)_{T_{c0}}}
=\frac{T_c}{T_{c0}}\frac{K_A^2}{K_B}.
\label{Cs}
\end{equation}

\section{Numerical Results.}

Numerical results for the case of charge (CDW) Ising -- like spin (SDW) 
fluctuations of short -- range (pseudogap) order were given in Ref. \cite{KSS}. 
Here we shall concentrate on the analysis of most important and interesting
case of Heisenberg spin (SDW) fluctuations and the role of impurity scattering
(disorder). Due to importance of $d$-wave pairing for copper oxide
superconductors, this case is analyzed in more details.

All calculations (in this section) were performed for typical values of
parameters of the ``bare'' spectrum $t'/t=-0.4$, $\mu /t=-1.3$, while for
the inverse correlation length we assumed $\kappa a=0.2$. To spare space
we do not present here explicit results for dimensionless GL -- coefficients
$K_A,\ K_B,\ K_C$, giving only the final results for the main physical
characteristics of a superconductor.

For dependences on the effective width of the pseudogap we always present
these characteristics normalized by the appropriate values at $T=T_{c0}$, 
while for the dependences on impurity scattering rate $\gamma_0$ we use
normalization by appropriate values at $T=T_{c0}(W)$, i.e. at the ``bare''
critical temperature for a given value of $W$, but in the absence of impurities
($\gamma_0=0$).

\subsection{$d$ -- wave pairing.}

\begin{figure}
\epsfxsize=14cm
\epsfysize=16cm
\epsfbox{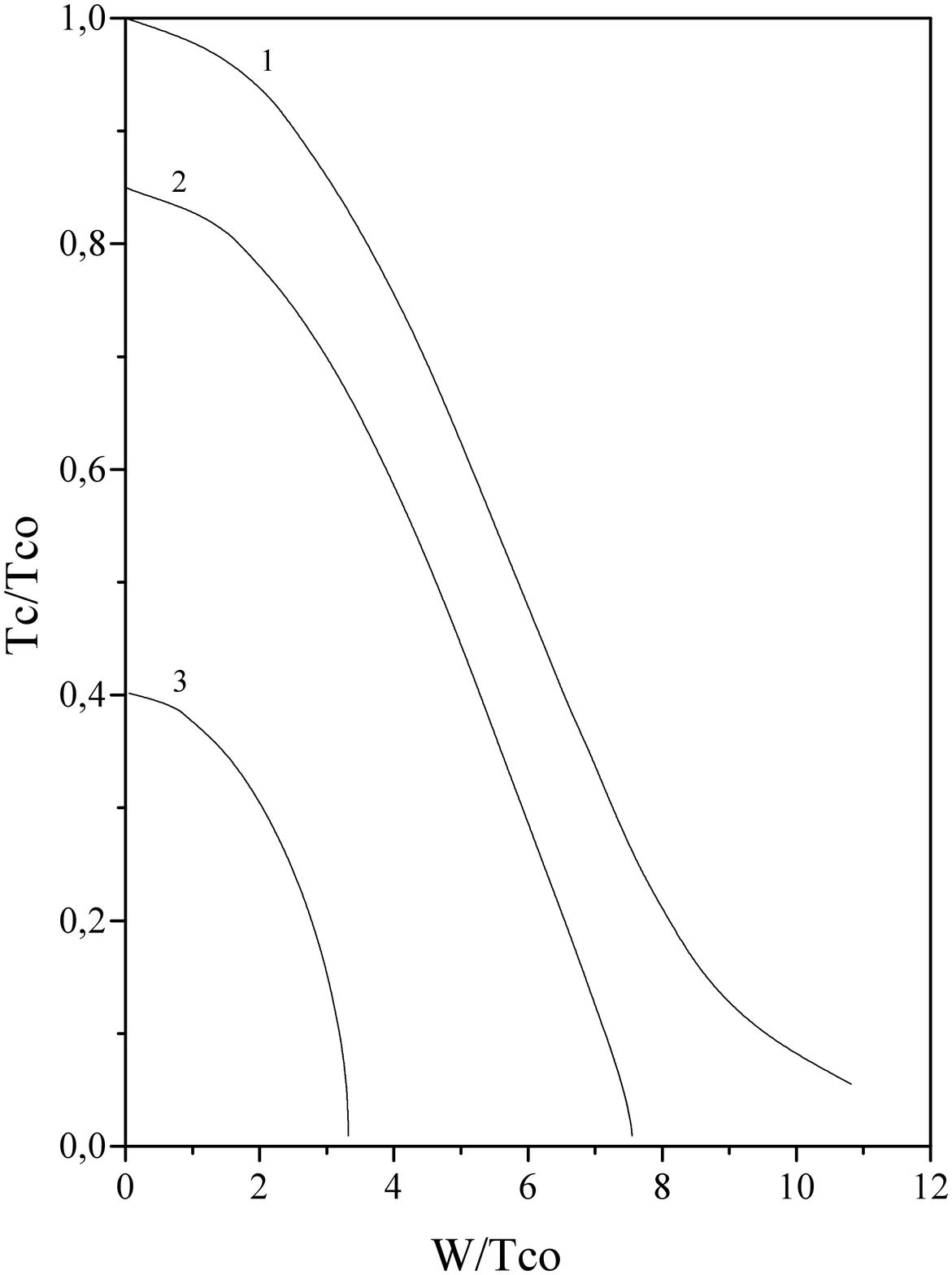}
\caption{$T_c$ dependence on the effective width of the pseudogap $W$ for
the case of $d$-wave pairing and for several values of impurity scattering
rate $\gamma_0/T_{c0}$: 0 -- 1; 0.18 -- 2; 0.64 -- 3.
Inverse correlation length of short -- range order $\kappa a$=0.2
} 
\label{sc1} 
\end{figure}

In Fig. \ref{sc1} we show the dependence of superconducting transition
temperature $T_c$ on the effective pseudogap width $W$ for several values
of impurity scattering rate. It can be seen that pseudogap fluctuations
lead to significant suppression of superconductivity and for finite disorder
we always obtain some ``critical'' value of $W$, when $T_c$ becomes zero.
This suppression of $T_c$ is naturally due to partial ``dielectrization''
of electronic spectrum in the vicinity of ``hot spots'' \cite{Sch,KS}.

\begin{figure}
\epsfxsize=14cm
\epsfysize=16cm
\epsfbox{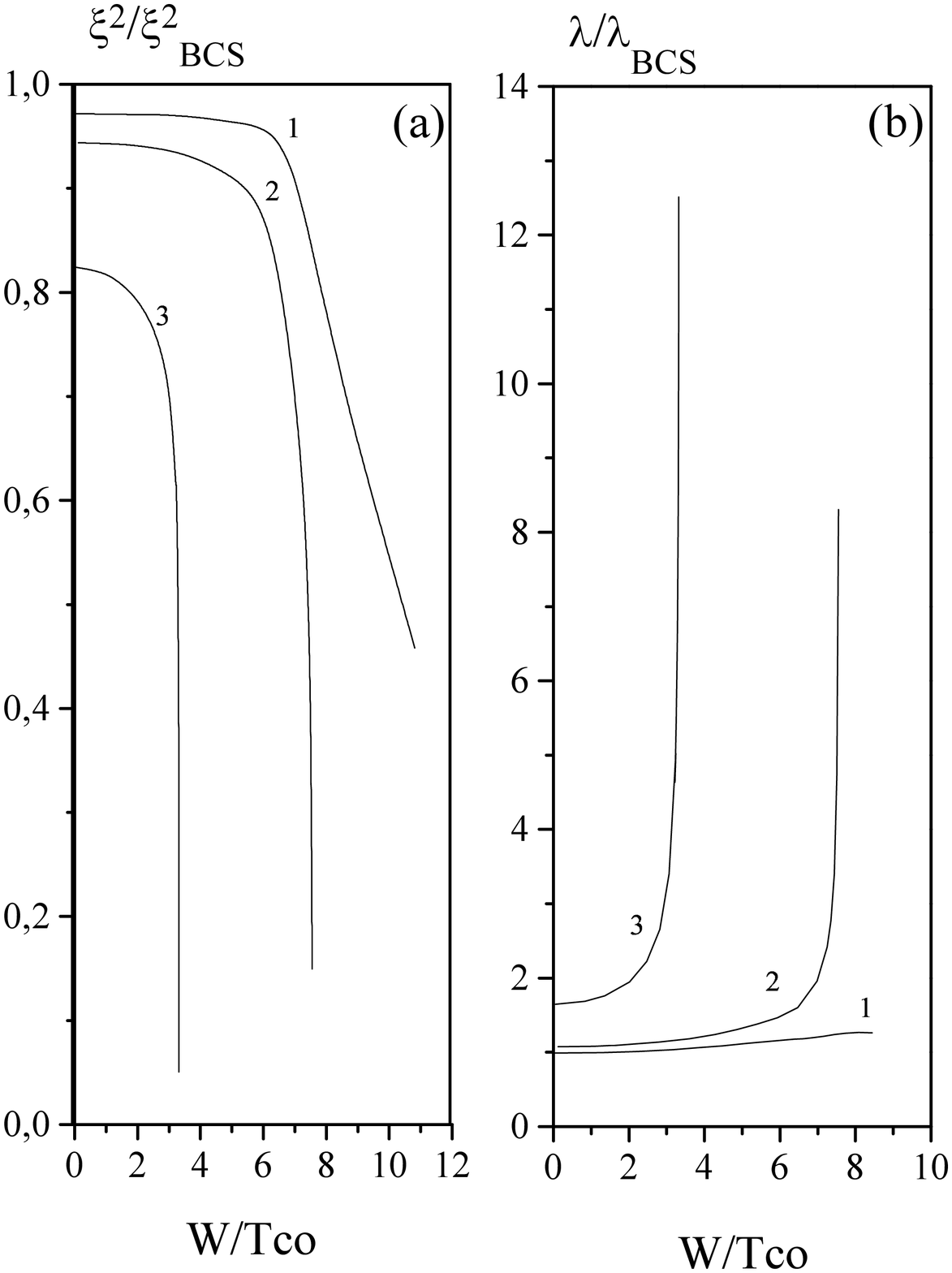}
\caption{Dependence of the square of coherence length and penetration depth 
on the effective width of the pseudogap $W$ for the case of $d$-wave pairing 
and for several values of impurity scattering rate $\gamma_0/T_{c0}$: 
0 -- 1; 0.18 -- 2; 0.64 -- 3.
} 
\label{sc2} 
\end{figure}

\begin{figure}
\epsfxsize=14cm
\epsfysize=16cm
\epsfbox{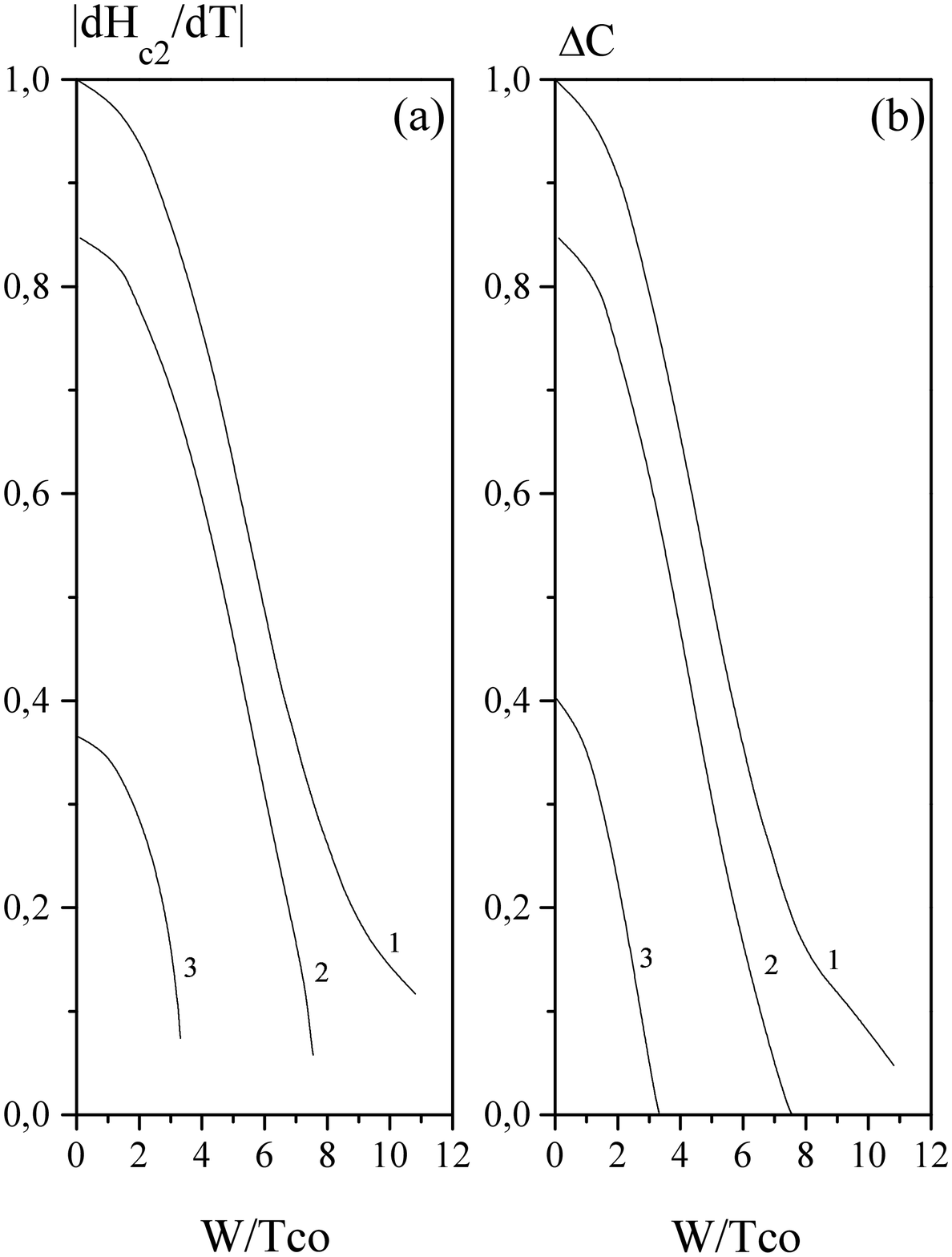}
\caption{Dependence of the slope of $H_{c2}(T)$ and specific heat discontinuity
at superconducting transition on the effective width of the pseudogap 
$W$ for the case of $d$-wave pairing and for several values of impurity 
scattering rate  $\gamma_0/T_{c0}$: 0 -- 1; 0.18 -- 2; 0.64 -- 3.
} 
\label{sc3} 
\end{figure}

Similar dependences for coherence length and penetration depth are shown in
Fig. \ref{sc2}, those for the slope of the upper critical field and specific
heat discontinuity at the transition  --- in Fig. \ref{sc3}. Typically we also
observe fast suppression of these characteristics by pseudogap fluctuations.

Dependences on the value of correlation length of short -- range order
fluctuations is, generally, more smooth --- in all cases the growth of
$\xi$ (drop of $\kappa$) enlarges the effects of pseudogap fluctuations.
We do not show appropriate results to spare space.

\begin{figure}
\epsfxsize=14cm
\epsfysize=16cm
\epsfbox{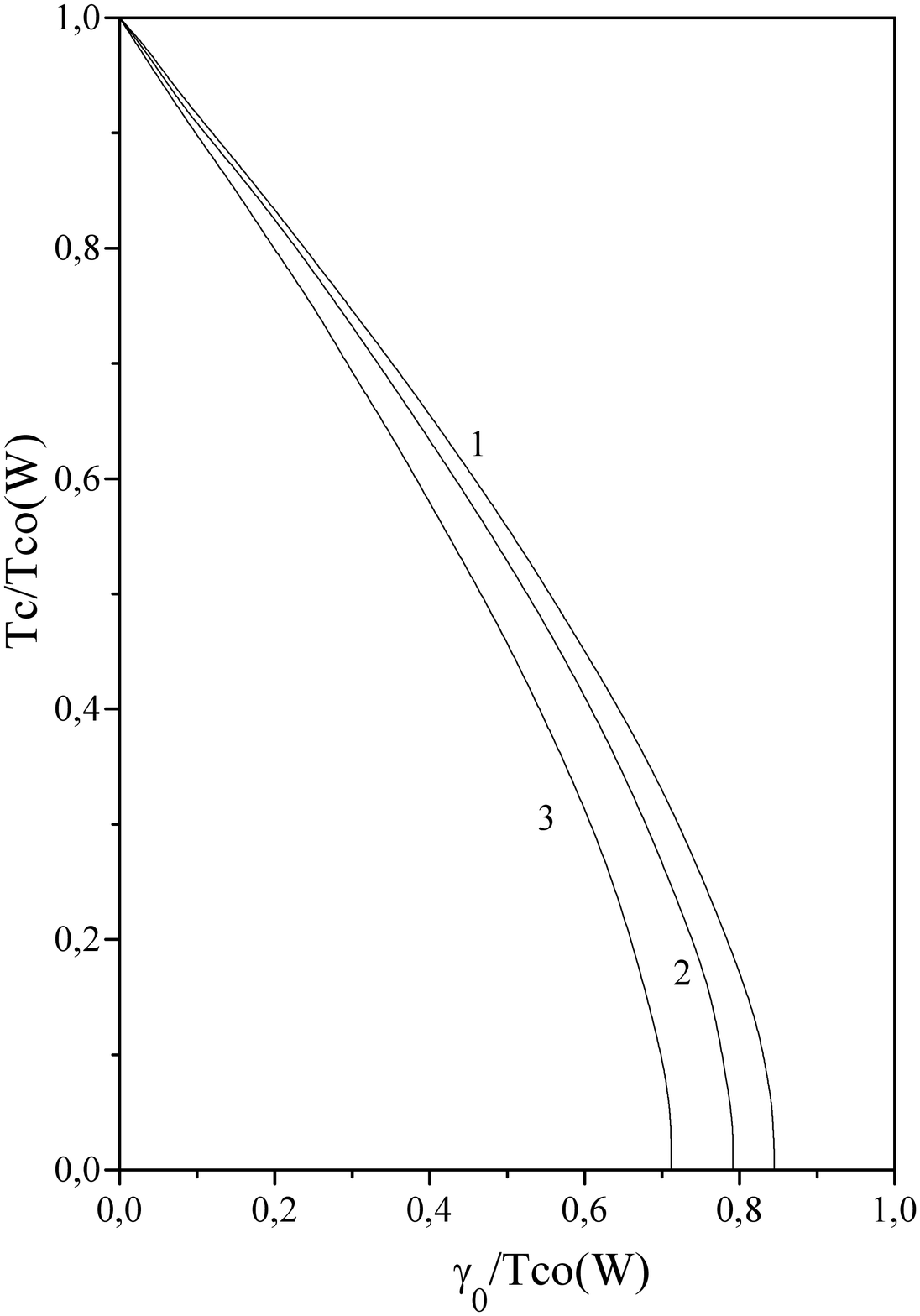}
\caption{$T_c$ dependence on the impurity scattering rate (disorder) 
$\gamma_0$ for the case of $d$-wave pairing and for several values of
the effective pseudogap width $W/T_{c0}$: 0 -- 1; 2.8 -- 2; 5.5 -- 3.
} 
\label{sc4} 
\end{figure}

\begin{figure}
\epsfxsize=14cm
\epsfysize=16cm
\epsfbox{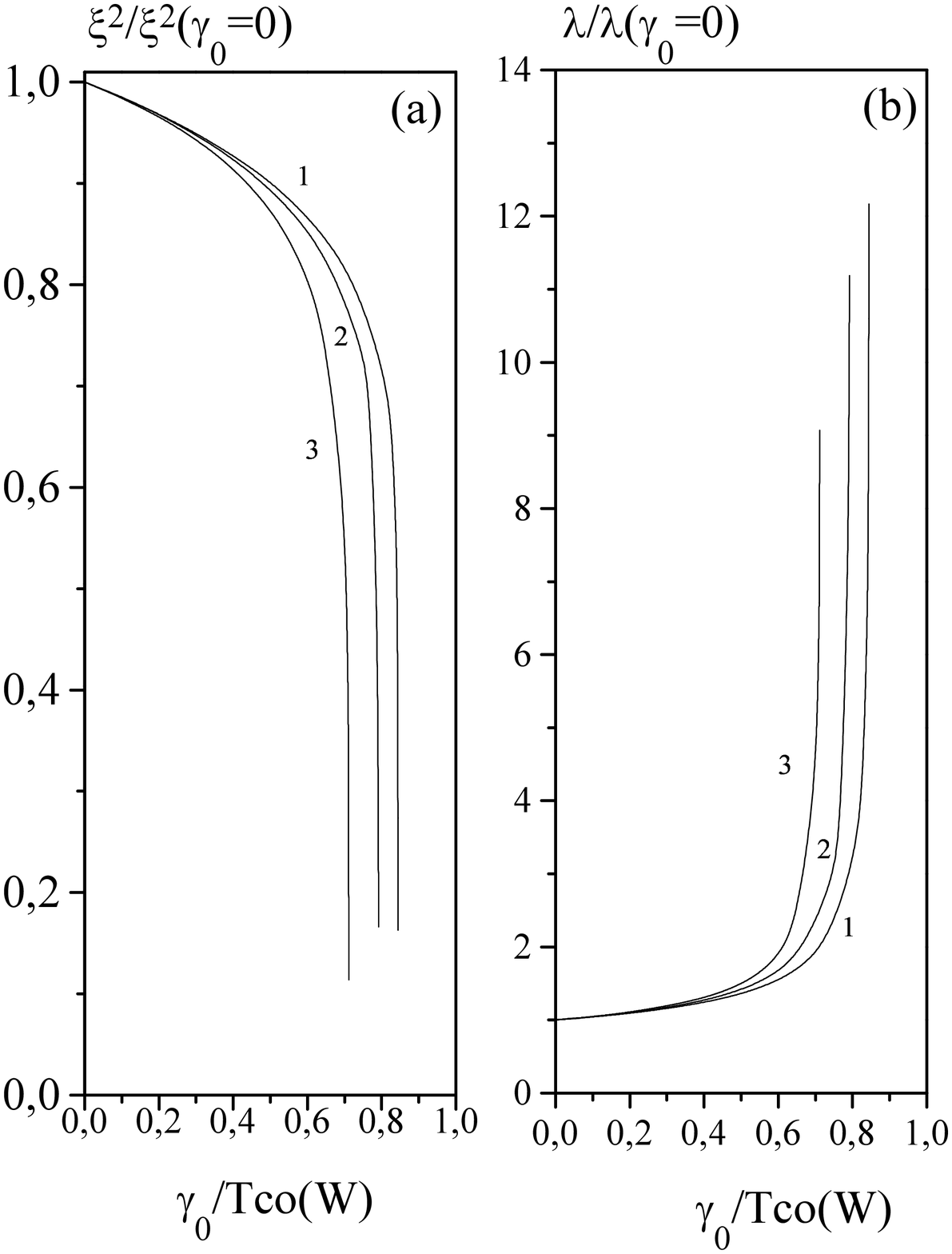}
\caption{Dependence of the square of the coherence length and penetration
depth on the impurity scattering rate (disorder) $\gamma_0$ for the case of
$d$-wave pairing and for several values of the effective pseudogap width 
$W/T_{c0}$: 0 -- 1; 2.8 -- 2; 5.5 -- 3.
} 
\label{sc5} 
\end{figure}

\begin{figure}
\epsfxsize=14cm
\epsfysize=16cm
\epsfbox{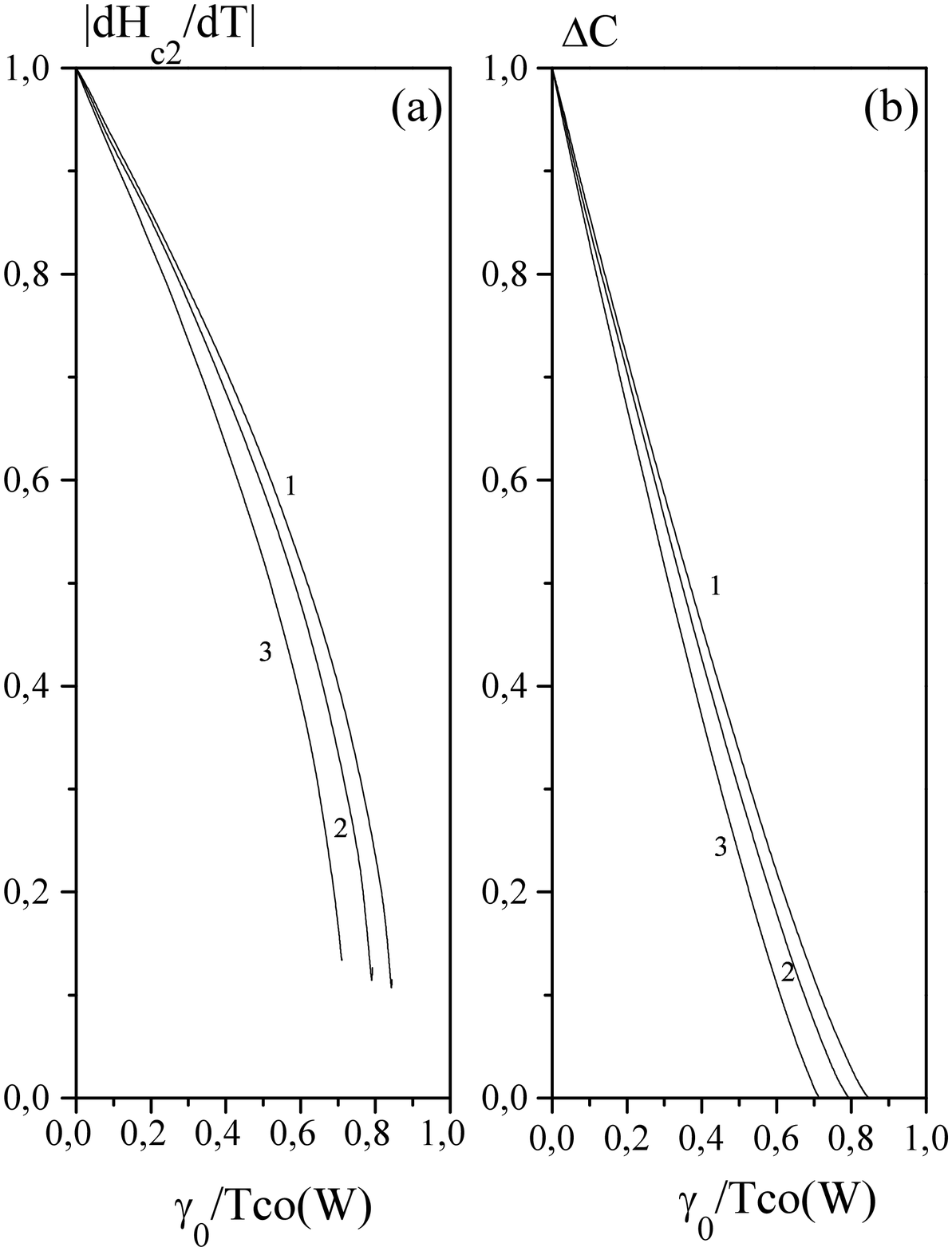}
\caption{Dependence of the slope of $H_{c2}(T)$ and specific heat discontinuity
at superconducting transition on the impurity scattering rate (disorder)
$\gamma_0$ for the case of $d$-wave pairing and for several values of 
the effective pseudogap width 
$W/T_{c0}$: 0 -- 1; 2.8 -- 2; 5.5 -- 3.
} 
\label{sc6} 
\end{figure}

In Fig. \ref{sc4} we show dependences of superconducting transition temperature
$T_c$ on impurity scattering rate $\gamma_0$ for several values of the 
effective pseudogap width. It is seen that in the presence of pseudogap
fluctuations $T_c$ suppression with the growth of disorder is faster than in
the absence of the pseudogap ($W=0$), when the dependence of $T_c$ on 
$\gamma_0$ in case of $d$-wave pairing, described by the standard Abrikosov --
Gorkov curve \cite{PS,Radt}. Analogous dependences for coherence length
and penetration depth are shown in Fig. \ref{sc5}, while for the slope of the
upper critical field $H_{c2}(T)$ and specific heat discontinuity --- in Fig.
\ref{sc6}. Again we see, that impurity scattering (disorder) leads to fast
suppression of the last two characteristics, e.g. the effect of pseudogap
fluctuations is enhanced by impurity scattering.

Dependences on parameters of the pseudogap, obtained above, are qualitatively
similar to those found in Ref. \cite{KSS} for the case of charged (CDW)
pseudogap fluctuations, when we also are dealing with recursion procedure for 
the vertex part with alternating signs. At the same time certain differences
appear here due another combinatorics of diagrams. Dependences on the
impurity scattering rate (disorder) in this model were not studied 
previously\footnote{Analysis of impurity dependence of $T_c$ in Ref. \cite{KK}
was performed for the case of recursion procedure without sign alternation,
appearing for the case of Ising -- like SDW fluctuations, when superconductivity
suppression is much slower.}.

These dependences are in qualitative agreement with majority of the data,
obtained in experiments on copper oxide superconductors in the pseudogap
region of the phase diagram (from underdoped to optimally doped region).
Below we shall demonstrate that these results may be used for direct
modeling of the typical phase diagram of these compounds.

\subsection{$s$ -- wave pairing.}

Analysis of the case of $s$-wave pairing is interesting mainly as demonstration
of characteristic differences with the case of $d$-wave pairing. At present,
there are no definite experimental data on $s$-wave superconductivity  in
systems with pseudogap, though it is quite possible that such systems will
be discovered in some future.

\begin{figure}
\epsfxsize=14cm
\epsfysize=16cm
\epsfbox{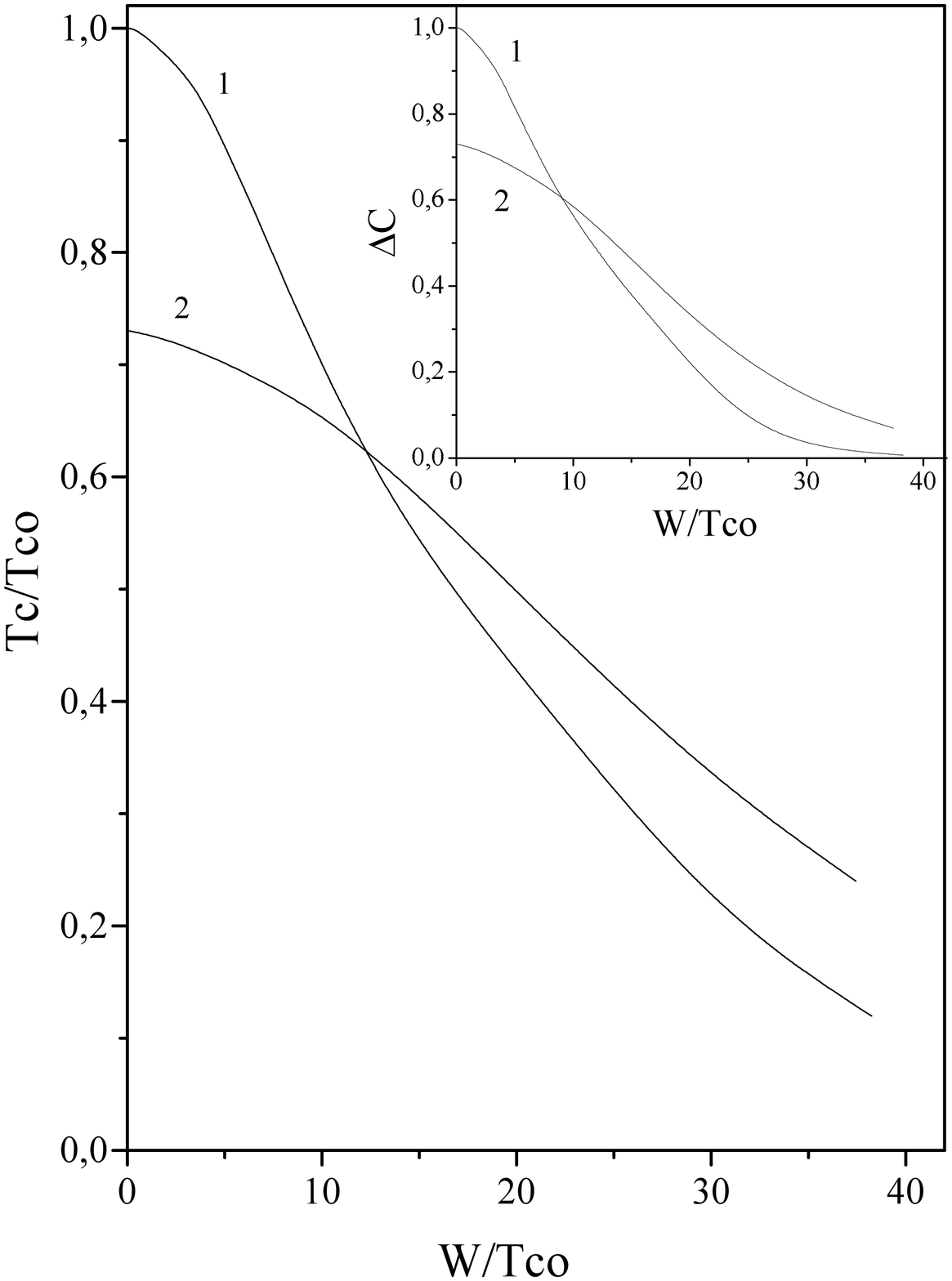}
\caption{$T_c$ dependence on the effective width of the pseudogap $W$ for
the case of $s$-wave pairing and for two values of the impurity scattering
rate $\gamma_0/T_{c0}$: 0 -- 1; 20 -- 2.
Inverse correlation length of short -- range order $\kappa a$=0.2.
At the insert --- characteristic dependence of specific heat discontinuity
for the same parameters.
} 
\label{sc7} 
\end{figure}

\begin{figure}
\epsfxsize=14cm
\epsfysize=16cm
\epsfbox{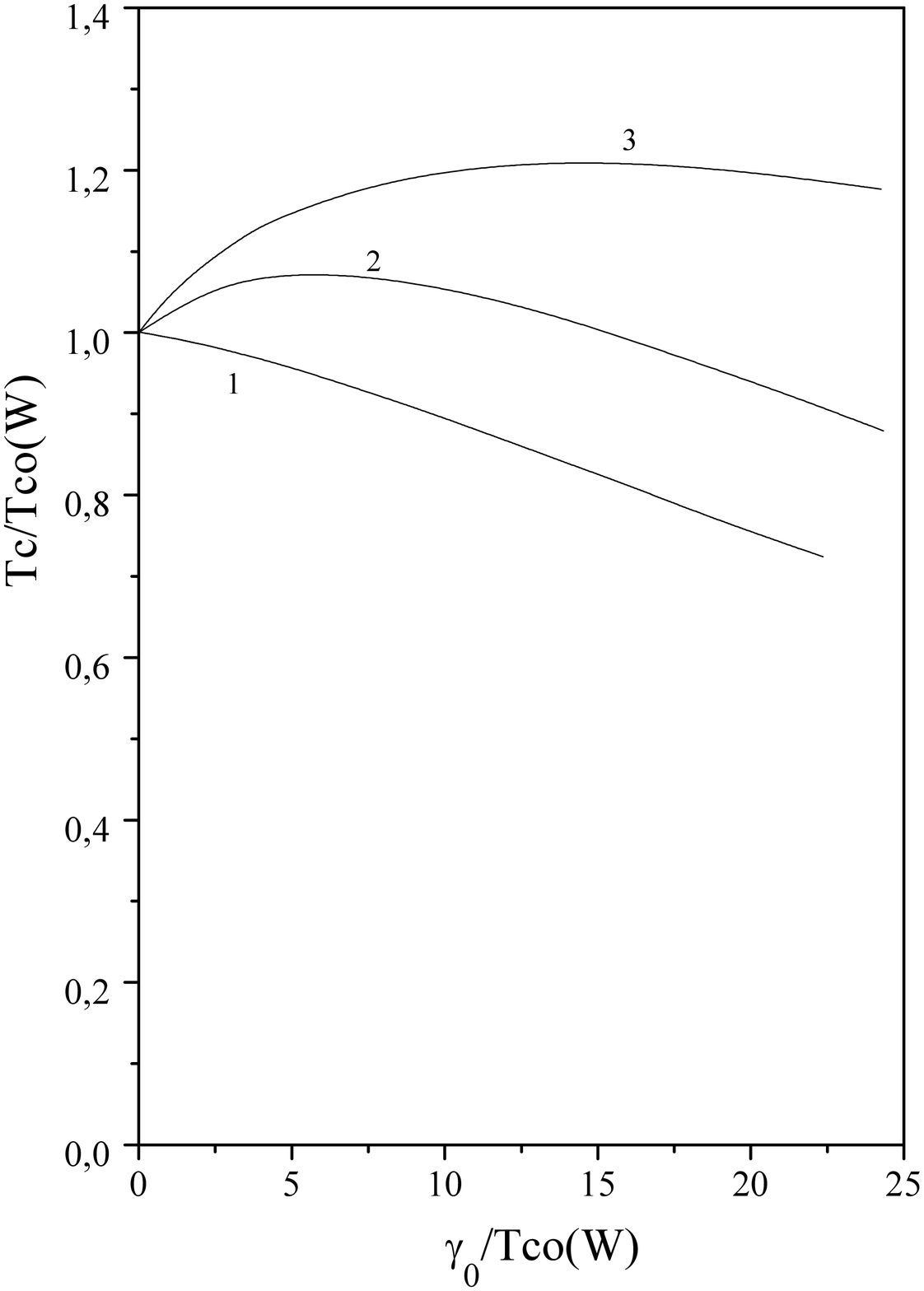}
\caption{$T_c$ dependence on the impurity scattering rate (disorder) 
$\gamma_0$ for the case of $s$-wave pairing and for several values of
effective pseudogap width $W/T_{c0}$: 0 -- 1; 8 -- 2; 15 -- 3.
} 
\label{sc9} 
\end{figure}

Our calculations show that pseudogap fluctuations significantly suppress
superconductivity also in this case (Fig. \ref{sc7}), though the characteristic
scale of these fluctuations, necessary for a major suppression of the 
superconducting state, is much larger here than in the case of $d$-wave pairing.
This result was obtained earlier in Ref. \cite{KSS}, though it is necessary
to note that in the case of Heisenberg (SDW) fluctuations, considered here,
there is no characteristic ``plateau'' in the dependence of $T_c$ on $W$, 
which was obtained for the case of charged (CDW) pseudogap fluctuations
in Ref. \cite{KSS}. On the same scale of $W$ takes place also the major
suppression of specific heat discontinuity at superconducting transition,
as shown at the insert in Fig. \ref{sc7}. Similar dependences for the
coherence length and penetration depth are analogous to those obtained in
Ref. \cite{KSS} and we do not show these here. In Fig.\ref{sc9} we show
dependences of $T_c$ on impurity scattering (disorder). Besides relatively
weak suppression of $T_c$ by disorder, connected \cite{KK} with small
suppression of the density of states at the Fermi level by disorder, we can 
observe also a weak growth of $T_c$ with $\gamma_0$, which is apparently due
some ``smearing'' of the pseudogap in the density of states.

\begin{figure}
\epsfxsize=14cm
\epsfysize=16cm
\epsfbox{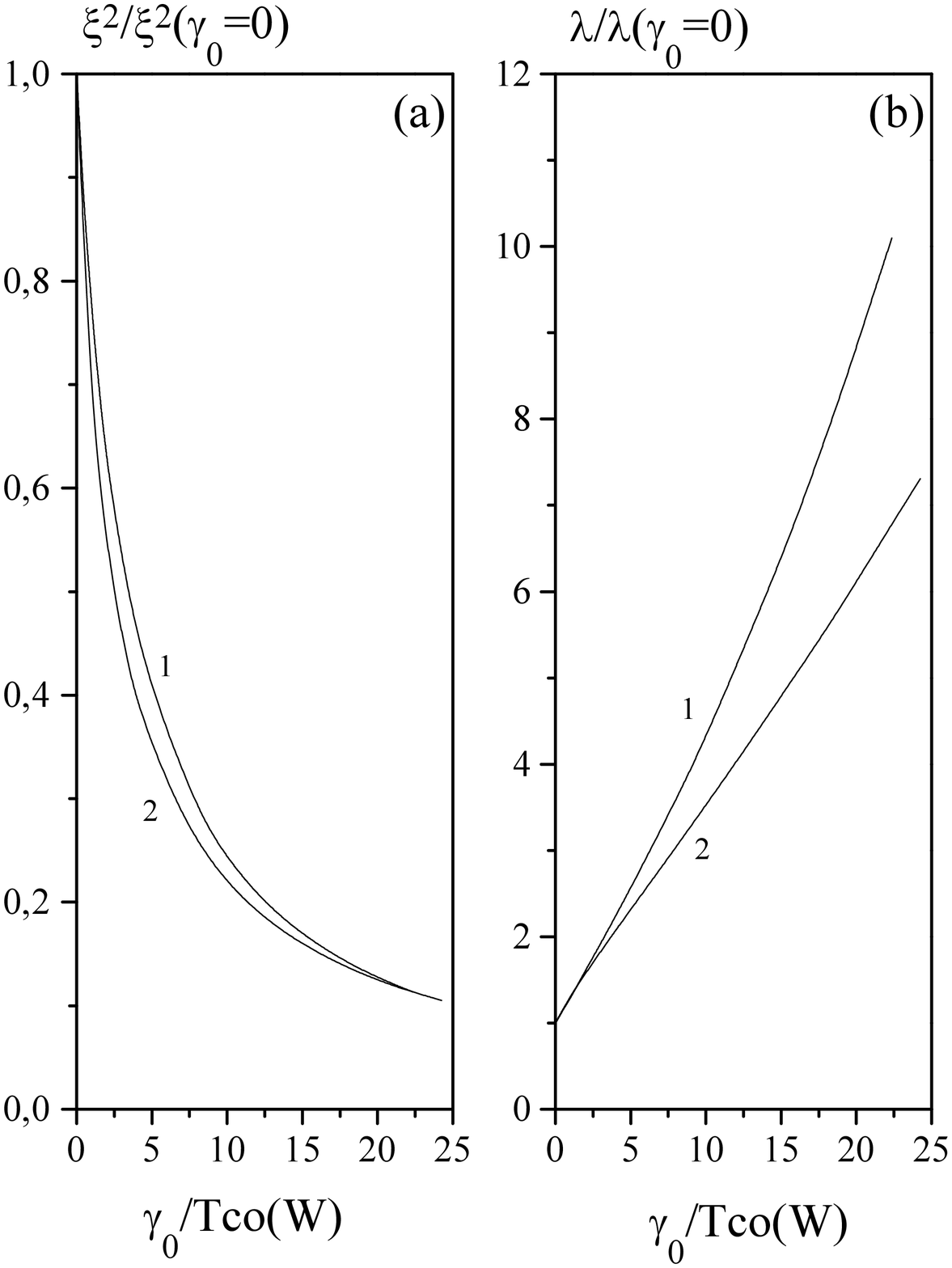}
\caption{Dependence of the square of coherence length and penetration depth
on the impurity scattering rate (disorder) $\gamma_0$ for the case of 
$s$-wave pairing for two values of the effective pseudogap width 
$W/T_{c0}$: 0 -- 1; 15 -- 2.
} 
\label{sc10} 
\end{figure}

In Fig. \ref{sc10} we show impurity effects for the case of $s$-wave
pairing on the coherence length and penetration depth.

\begin{figure}
\epsfxsize=14cm
\epsfysize=16cm
\epsfbox{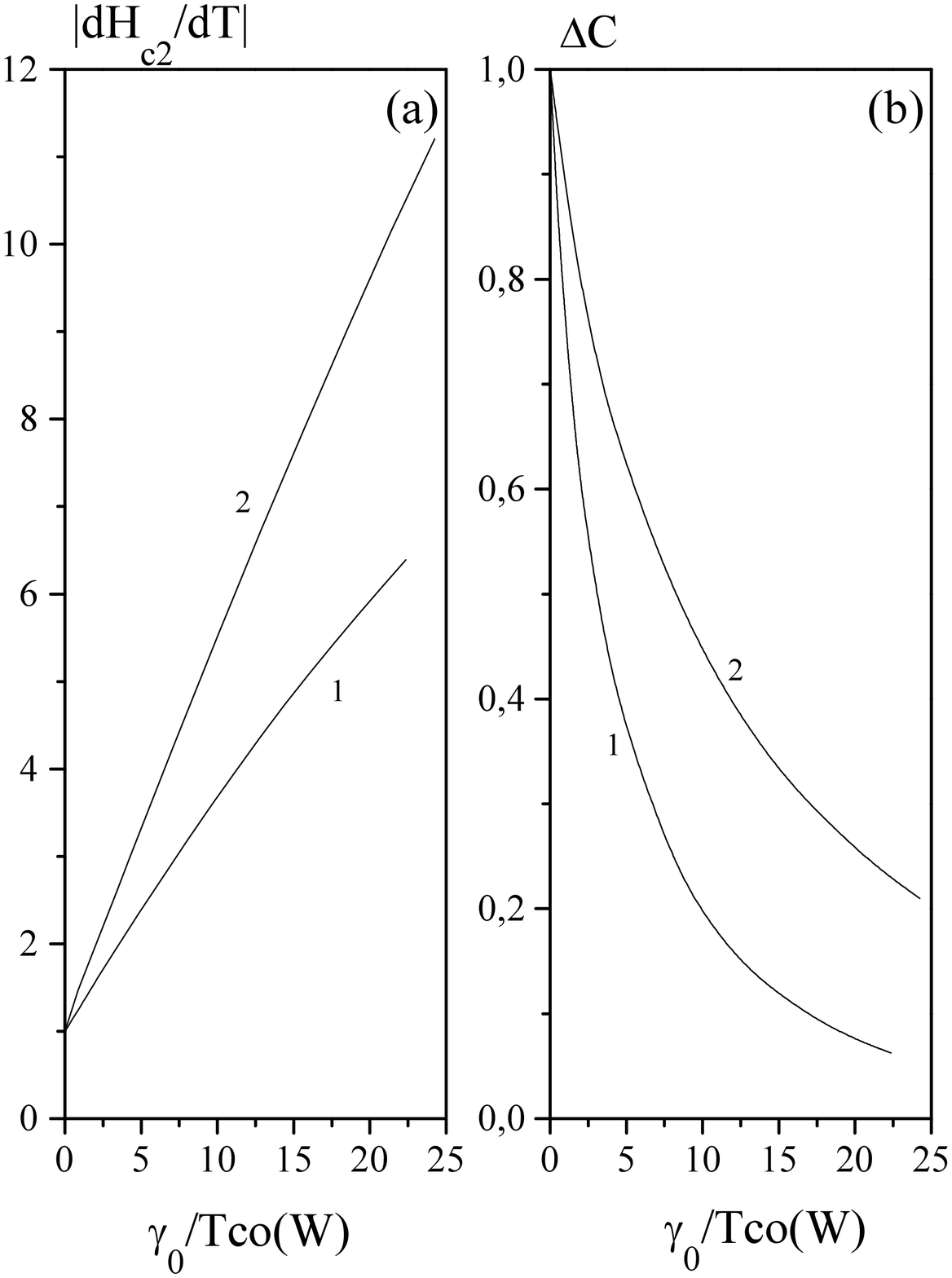}
\caption{Dependence of the slope of $H_{c2}(T)$ and specific heat discontinuity
at the transition on the impurity scattering rate (disorder) $\gamma_0$ for
the case of $s$-wave pairing and for two values of the effective pseudogap
width $W/T_{c0}$: 0 -- 1; 15 -- 2.
} 
\label{sc11} 
\end{figure}

Finally, in Fig. \ref{sc11} we demonstrate the influence of impurity scattering
(disorder) on the slope of the upper critical field and specific heat 
discontinuity. Again the specific heat discontinuity is significantly
suppressed by disorder, while the slope of $H_{c2}(T)$ demonstrates
qualitatively different behavior, than in the case of the $d$-wave pairing: 
the growth of disorder leads to the growth of this slope, as in the standard
theory of ``dirty'' superconductors \cite{Scloc}, while pseudogap
fluctuations further increase the slope of $H_{c2}(T)$. In the absence of
pseudogap fluctuations, similar differences (between $s$-wave and $d$-wave
cases) in  the behavior of the slope of $H_{c2}(T)$ under disordering were 
noted earlier in Ref. \cite{PS}.

\section{Modeling of the Phase Diagram.}

The described above model of the influence of pseudogap fluctuations on
superconductivity allows to perform simple modeling of a typical phase
diagram of high -- temperature superconducting cuprates\footnote{We ``neglect''
here the existence of a narrow region of antiferromagnetic ordering in Mott
dielectric appearing at low concentrations of doping impurity, limiting our
analysis to a wide region of a ``bad'' metal.}. First attempt of such modeling
in an oversimplified version of our model was made in Ref. \cite{AC}. 
The main idea is in identifying our parameter $W$ with an experimentally
observable effective width of the pseudogap (crossover temperature into
pseudogap region of the phase diagram) $E_g\approx T^*$, which is determined
by numerous experiments \cite{Lor,MS,Pines}. It is well known that this 
crossover temperature is practically linear in concentration of doping 
impurity (carrier concentration), starting at small concentration from the
values of the order of $10^3$K and going to zero at some critical concentration
$x_c\approx 0.19..0.22$, which is slightly higher than the ``optimal'' value 
$x_o\approx 0.15..0.17$ \cite{Lor,NT}. Accordingly, we can assume similar
concentration dependence of our pseudogap parameter $W(x)$\footnote{Obviously,
this identification can be done up to some unknown factor $\sim 1$.}. 
In this sense we can say that concentration dependence $W(x)$ directly follows
from the experiment. Then the only parameter of the model to be determined
remains the concentration dependence of the ``bare'' superconducting
transition temperature $T_{c0}(x)$, which would have existed in the absence 
of pseudogap fluctuations. The knowledge of this dependence allows then to
calculate the observed concentration dependence of superconducting transition
temperature $T_c(x)$ solving equations of our model. Unfortunately, as was
already noted in Ref. \cite{KSS}, concentration dependence $T_{c0}(x)$ is, in
general case, unknown and is not determined from any known experiment.
Thus, it remains the fitting parameter of our theory.

\begin{figure}
\epsfxsize=15cm
\epsfysize=17cm
\epsfbox{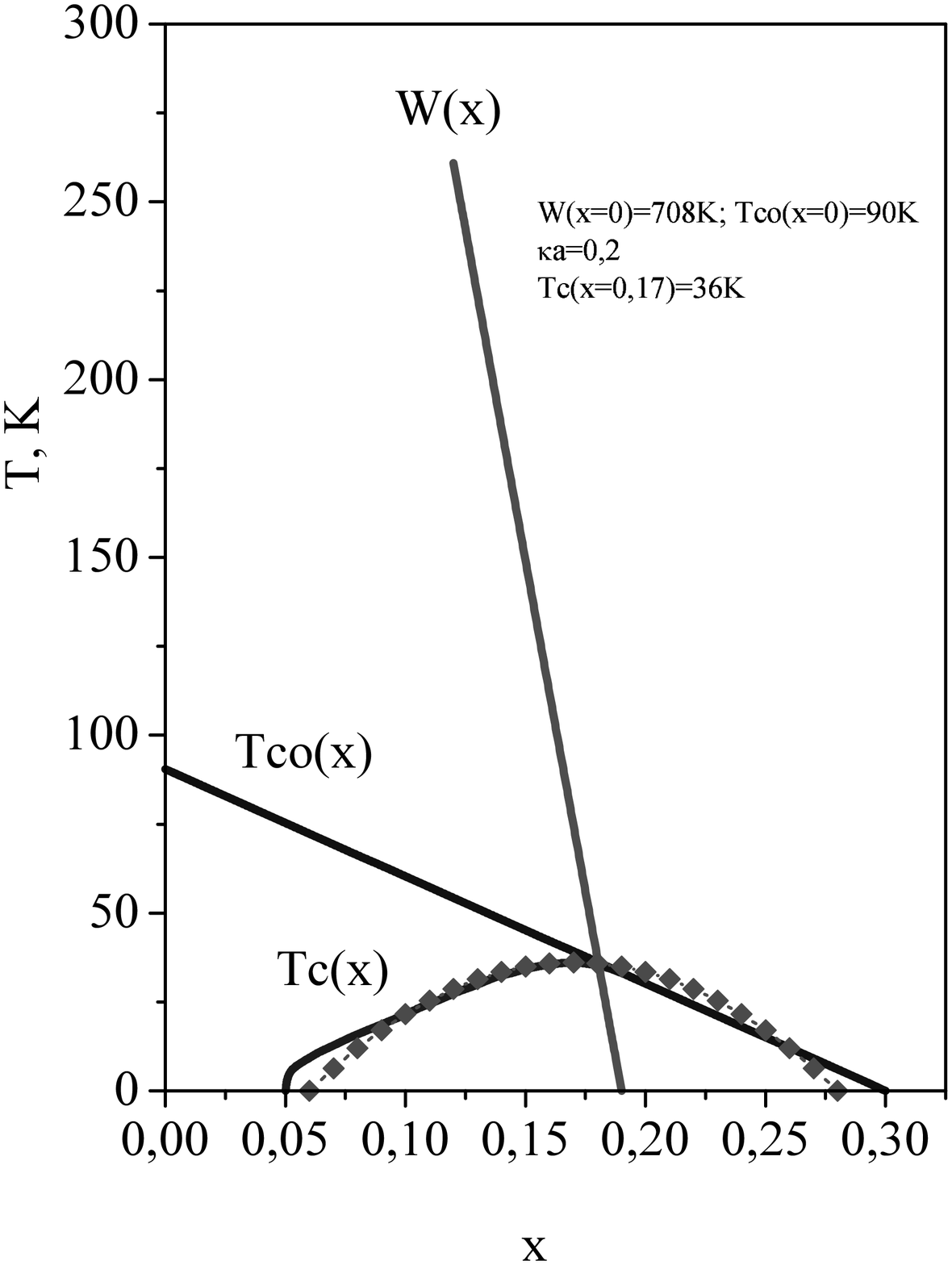}
\caption{Model phase diagram for the case of pseudogap fluctuations of
CDW type ($d$-wave pairing) and ``bare'' superconducting transition temperature
$T_{c0}$ with linear dependence on carrier concentration.}  
\label{TcCDW} 
\end{figure}

Assuming following Ref. \cite{AC} that $T_{c0}(x)$ can also be described
by a linear function of concentration $x$, going to zero at $x=0.3$, 
and varying the value of $T_{c0}(x=0)$ to obtain the correct value of
$T_c(x=x_o)$ it is possible to calculate the ``observed'' dependence of
$T_c(x)$. An example of such calculation for the case of 
$d$-wave pairing and scattering by charged (CDW) pseudogap fluctuations 
\cite{KSS}, using typical dependence of $W(x)$, is shown in Fig. \ref{TcCDW}.  
It can be seen that even with such crude and rather arbitrary assumptions,
the ``hot spots'' model allows to obtain the $T_c(x)$ dependence, which is
pretty close to the experimentally observed. Similar calculations for the
case of Ising -- like model of interaction of electrons with spin fluctuations
(no sign alternation in the recursion procedure for the vertex \cite{KSS}) 
show that the reasonable values of $T_c(x)$ can only be obtained for rather
unrealistic values of $W(x)$ about an order of magnitude higher than those 
observed experimentally.

\begin{figure}
\epsfxsize=16cm
\epsfysize=18cm
\epsfbox{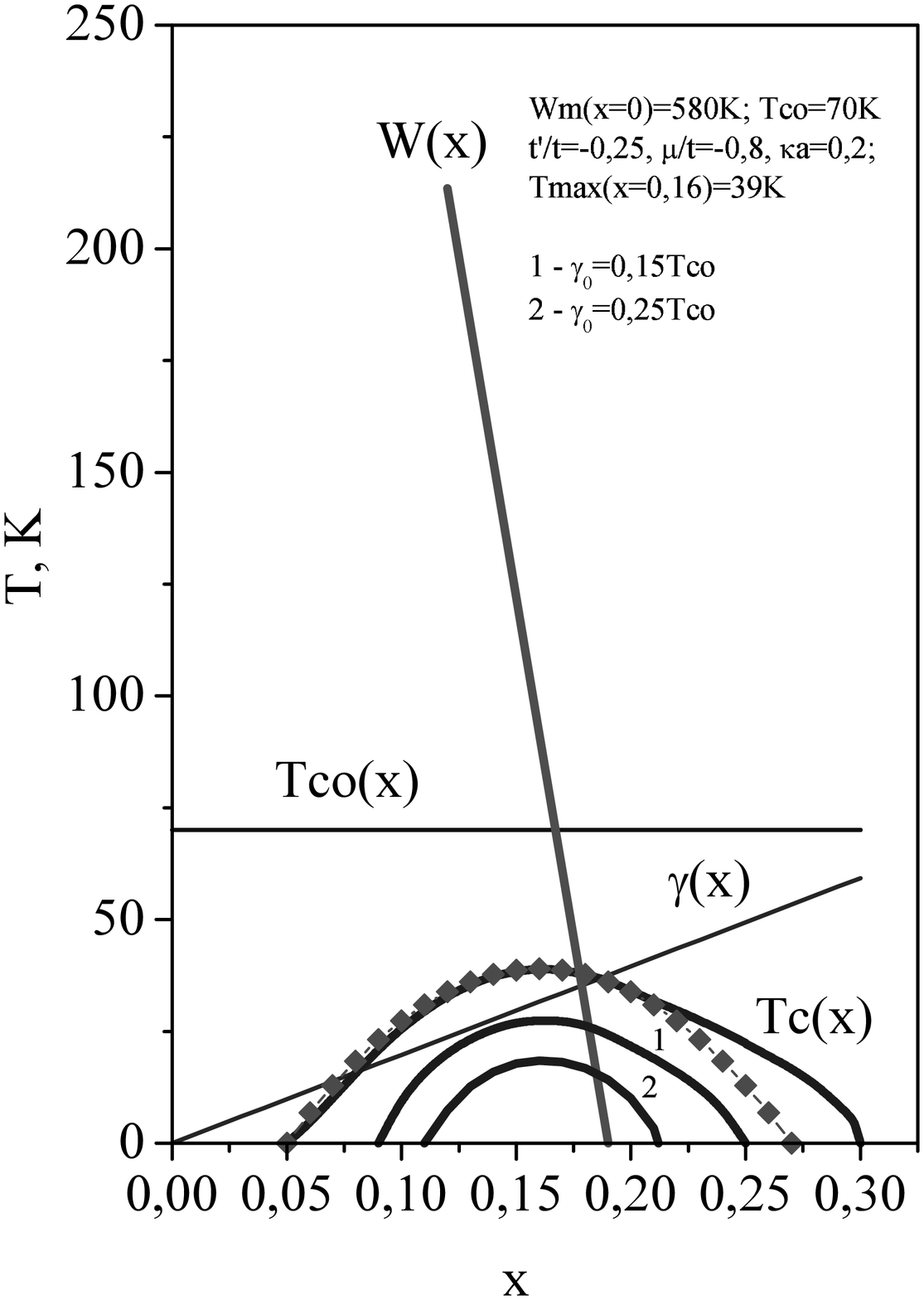}
\caption{Model phase diagram for the case of Heisenberg (SDW) pseudogap
fluctuations ($d$-wave pairing) and ``bare'' superconducting transition
temperature $T_{c0}$ independent of carrier concentration, taking into
account the role of internal disorder, linear over the concentration of
doping impurity $\gamma(x)$.}  
\label{TcSDW} 
\end{figure}

Within our BCS -- like model for the ``bare'' $T_{c0}$, an assumption of
significant concentration dependence of this parameter seems rather
unrealistic\footnote{Within this approach any dependence of $T_{c0}$ on $x$ 
can only be due to some weak concentration dependence of the density of
states on the Fermi level.}. Thus, let us assume that the value of $T_{c0}$ 
has no dependence on carrier (doping impurity) concentration $x$ at all, but
take into account the fact that introduction of doping impurity immediately
leads to the appearance of some random impurity scattering of carriers
(due to internal disorder), which can be simply described by introduction of
some linearly growing impurity scattering rate $\gamma(x)$. Now let us just
assume that this growth of internal disorder leads to complete suppression
of $d$-wave pairing at $x=0.3$, in accordance with Abrikosov -- Gorkov
dependence \cite{PS,Radt}. Results of calculations for the phase diagram of
$La_{2-x}Sr_xCuO_4$ -- type system within this model, and for the case of
Heisenberg pseudogap fluctuations and the described role of impurity
scattering is shown in Fig. \ref{TcSDW}. The values of different parameters
used in this calculation are shown at the same figure. ``Experimental'' values
of $T_c(x)$, shown in this figure (as well as in Fig. \ref{TcCDW}) by
``diamonds'', were obtained from the empirical relation \cite{NT,PT}:
\begin{equation}
\frac{T_c(x)}{T_c(x=x_o)}=1-82.6(x-x_o)^2
\label{Tcexp}
\end{equation}
which represents rather good fit to experimental data for concentration
dependence of $T_c$ in a number of cuprates. We can see that in all
underdoped region our model gives practically ideal description of
``experimental'' data for quite reasonable values of $W(x)$. At the end of
overdoped region agreement is less satisfactory, but it should be noted that 
both the relation (\ref{Tcexp}) is not quite good here, and our model of
superconductivity suppression in overdoped region is obviously too crude.
Also it should be noted that we have not performed any special fitting of 
parameters of our model to increase agreement in this region. 

It is interesting to analyze the behavior of superconducting transition
temperature $T_c$ under additional disordering at different compositions
(concentrations of carriers). There were numerous experimental studies with
such disordering being introduced by additional impurities \cite{Uch,Tall} 
or by irradiation by fast neutrons \cite{Gosch} and electrons \cite{Tolp,Rull}. 
Special discussion of the role of such additional disordering in the context
of the existence of the pseudogap state was presented, apparently, only
in Ref. \cite{Tall}.

In our model such additional disordering is easily simulated by the
introduction of an additional impurity scattering rate $\gamma_0$, which is
just added to our parameter of an internal disorder $\gamma(x)$.
Results of our calculations for two values of this parameter are also shown
in Fig. \ref{TcSDW}. We can see, that in complete agreement with experiments
\cite{Tall}, introduction of additional ``impurities'' (disorder) leads to
rather fast reduction of the superconductivity region on the phase diagram
Also in complete agreement with our conclusion made above with respect to
Fig. \ref{sc4} and with the results of experiments \cite{Gosch,Tall}, 
superconductivity suppression by disordering in underdoped (pseudogap)
region is significantly faster, than for optimal doping. It could be expected
that introduction of ``normal'' disorder, obviously leading to some
suppression of the pseudogap in the density of states, could have lead to some
``slowing down'' of $T_c$ suppression. However, our calculations show that
such effect in the case of $d$-wave pairing is just absent. 

The problem, however, is that in all cases our calculations show that $T_c$
suppression is faster than it is predicted by the standard Abrikosov -- 
Gorkov curve in case of $d$-wave pairing \cite{PS,Radt}. At the same time,
most attempts to fit experimental data on disordering in cuprates
\cite{Uch,Tolp,Rull}, show that such suppression is in fact significantly
{\em slower}, than predicted by Abrikosov -- Gorkov dependence. This fact
remains one of the major unsolved problems of the theory of high --
temperature superconductors \cite{Scloc}. One of the possible solutions of this
problem may be connected with rigorous description of disordering effects in
superconductors, which belong to transition region from ``large'' pairs of
BCS theory to ``compact'' pairs (bosons), appearing in the limit of very strong
coupling \cite{PosSad}. Another interesting possibility to explain the 
``slowing down'' of $T_c$ suppression is connected with an account of
anisotropy of impurity scattering, analyzed in detail in Refs. \cite{PS,HN}. 
This effect can be rather easily included in our calculational scheme.
It seems especially significant in connection with the established strong
anisotropy of elastic scattering (with $d$-wave symmetry), observed in ARPES
experiments on $Bi_2Sr_2CaCu_2O_{8+\delta}$ \cite{Valla,Kam}. Appropriate
scattering rate changes in the interval of $20-60 meV$ \cite{Kam}, which is
nearly an order of magnitude higher, than the maximal value of $\gamma(x)$,
used in our calculations, which once again demonstrates an unusual stability
of $d$-wave pairing in cuprates towards static disorder. It should be noted
that our model of electron self -- energy in fact describes analogous 
anisotropy of elastic scattering, corresponding to its growth in the vicinity
of ``hot spots'', but we do not observe the effect of the ``slowing down''
of $T_c$ suppression in our calculations.

Our results show that despite the obvious crudeness of our assumptions, the
``hot spots'' model allows to obtain rather reasonable (even semiquatitative)
description of the region of existence of superconducting state on the phase
diagram of high -- temperature superconducting cuprates\footnote{Above we
always assumed that we are dealing with hole -- doped cuprates, where
concentration dependence of $T^*(x)$ is well established \cite{Lor,NT}.  
For electron doped systems the data on the pseudogap state are rather
fragmentary.}. The main deficiency of our approach remains rather indeterminate
scenario for concentration dependence of the ``bare'' superconducting
transition temperature.

\section{Conclusion}

The analysis presented above shows, that the model of the pseudogap state,
based on the concept of ``hot spots'', can provide a reasonable description
of the main properties of the superconducting phase of cuprates, as well as
of the phase diagram, using relatively small number of fitting parameters,
most of which can be determined from independent experiments.

Let us stress that all calculations were performed using the standard assumption
of self -- averaging nature of superconducting order parameter (gap) over the
random fields of impurities and pseudogap fluctuations. Usually this 
assumption is justified for superconductors with coherence lengths (size of
Cooper pairs) much exceeding the other microscopic lengths in the system,
such as mean free path or correlation length of pseudogap fluctuations $\xi$. 

In the class of models of the pseudogap state under consideration this 
assumption is not, in general, valid and important effects due to non self --
averaging \cite{KS01,KS02}, which lead to a qualitatively new picture of
inhomogeneous superconducting state, with ``drops'' of superconducting phase 
existing for temperatures $T>T_{c}$. In principle, there are experimental data,
directly supporting such picture of inhomogeneous superconductivity in
cuprates \cite{Pan,McE,Kap}. Of course, we are far from claiming, that these
experiments directly support oversimplified theoretical models developed in
Refs. \cite{KS01,KS02}. At the same time, these results stress the 
importance of rigorous analysis of non self -- averaging effects in relatively
``realistic'' models of the pseudogap state, such as the ``hot spots'' model,
described above\footnote{The picture of  ``superconducting drops'', existing
above $T_c$ allows to understand a number of experiments, which are usually
interpreted as supporting the superconducting nature of the pseudogap in
cuprates.}. 

This work was supported in part by the RFBR grant No. 02-02-16031, as well as
by the program of fundamental research ``Quantum macrophysics'' of the 
Presidium of RAS, the program ``Strongly correlated electrons in 
semiconductors, metals, superconductors and magnetic materials'' of the 
RAS Department of Physical Sciences, and also under the project of the
Ministry of Education and Science of the Russian Federation
``Studies of collective and quantum effects in condensed matter''.

\newpage

\appendix

\section{Diagram Combinatorics for Heisenberg Pseudogap Fluctuations.}

To analyze diagram combinatorics, let us consider the limit of infinite
correlation length of pseudogap fluctuations. In this case the spin density
scattering electrons takes the form:
\begin{equation}
{\bf S_q}={\bf S}\delta ({\bf q}-{\bf Q})
\label{Spden}
\end{equation}
and averaging over Gaussian spin fluctuations reduces to the usual integration
\cite{Sch}:
\begin{equation}
<\ldots>=\frac{g^3}{(2\pi )^{\frac{3}{2}}W^3}
\int d{\bf S}e^{-\frac{g^2{\bf S}^2}{2W^2}}\ldots
\label{SpFl1}
\end{equation}
Accordingly, in this limit we can first solve the problem for an electron,
moving in the coherent field of spin density (\ref{Spden}), and afterwards
perform averaging (\ref{SpFl1}) over its fluctuations. For further analysis
it is convenient to introduce fluctuating field
${\bf\delta}=\frac{g}{\sqrt{3}}{\bf S}$ --- ``potential'', scattering electrons.
Then averaging (\ref{SpFl1}) over spin fluctuations reduces to the averaging 
over fluctuations of this random field:
\begin{equation}
<\ldots>=\left(\frac{3}{2\pi W^2}\right)^{\frac{1}{2}}
\int_{-\infty}^{+\infty} d\delta _le^{-\frac{3\delta _l^2}{2W^2}}
\frac{3}{2\pi W^2}\int_{0}^{2\pi} d\varphi
\int_{0}^{+\infty}d|\delta _t||\delta _t|e^{-\frac{3|\delta _t|^2}{2W^2}}
\ldots
\label{SpFl2}
\end{equation}
We see that actually we have two fluctuating fields scattering electrons:
the real longitudinal field $\delta _l=\frac{g}{\sqrt{3}}S_z$ and complex
transverse $\delta _t$, characterized by the amplitude $|\delta _t|$ and 
phase $\varphi$, connected with two transverse components of $\bf S$.

Such averaging generates diagram technique with two types of effective 
interactions \cite{Sch}. One can be represented say by dashed interaction
line and correspond to: 
\begin{equation}
V_{eff1}=\frac{g^2}{3}<S_{z{\bf q}}S_{z-{\bf q}}>=
\pm\frac{W^2}{3}\delta({\bf q}-{\bf Q})
\label{Veff1}
\end{equation}
where the sign ``$-$'' is taken when spin projection under this line is
changed (e.g. when this dashed line surrounds an odd number of spin --
flipping operators $S_+$, $S_-$), and another, represented by wavy line:
\begin{equation}
V_{eff2}=\frac{g^2}{3}<S_{+{\bf q}}S_{--{\bf q}}>=
2\frac{W^2}{3}\delta({\bf q}-{\bf Q})
\label{Veff2}
\end{equation}
The averages like $<S_+S_+>$ and $<S_-S_->$ are zero after averaging over the 
phase in (\ref{SpFl2}).

Let us start with electron moving in coherent filed of spin density
(\ref{Spden}). In this case single -- particle Green's function is represented
by $2\times 2$ matrix with 4 independent components\footnote{Components 
differing from these by the change of all spin projections can be obtained 
by substitution: 
$\delta _l\to -\delta _l$, $\delta _t\leftrightarrow\delta ^*_t$}, which are
determined by the following system:
\begin{eqnarray}
G_{1\uparrow ;1\uparrow}&=&G_1+G_1\delta _lG_{2\uparrow ;1\uparrow}
+G_1\delta _tG_{2\downarrow ;1\uparrow}\nonumber\\
G_{2\uparrow ;1\uparrow}&=&G_2\delta _lG_{1\uparrow ;1\uparrow}
+G_2\delta _tG_{1\downarrow ;1\uparrow}\nonumber\\
G_{2\downarrow ;1\uparrow}&=&-G_2\delta _lG_{1\downarrow ;1\uparrow}
+G_2\delta ^*_tG_{1\uparrow ;1\uparrow}\nonumber\\
G_{1\downarrow ;1\uparrow}&=&-G_1\delta _lG_{2\downarrow ;1\uparrow}
+G_1\delta ^*_tG_{2\uparrow ;1\uparrow}
\label{systG}
\end{eqnarray}
where we have introduced short notations:\ 
$(\varepsilon_n,{\bf p})\to 1$,
$(\varepsilon_n,{\bf p}+{\bf Q})\to 2$ and 
$G_1=\frac{1}{i\varepsilon_n-\xi_{\bf p}}$, 
$G_2=\frac{1}{i\varepsilon_n-\xi_{{\bf p}+{\bf Q}}}$. Then we obtain:
\begin{eqnarray}
G_{1\uparrow ;1\uparrow}=\frac{G_2^{-1}}{G_1^{-1}G_2^{-1}-|\delta |^2};
&\qquad &
G_{2\uparrow ;1\uparrow}=\frac{\delta _l}{G_1^{-1}G_2^{-1}-|\delta |^2}
\nonumber\\
G_{1\downarrow ;1\uparrow}=0; &\qquad &
G_{2\downarrow ;1\uparrow}=\frac{\delta ^*_t}{G_1^{-1}G_2^{-1}-|\delta |^2}
\label{matG}
\end{eqnarray}
where $|\delta |=\sqrt{\delta_l^2+|\delta_t|^2}$ is the amplitude of the
field ${\bf\delta}$.

The averaged Green's function is given now by:
\begin{equation}
G=<G_{1\uparrow ;1\uparrow}>=
\sqrt{\frac{2}{\pi}}\frac{1}{(W^2/3)^{\frac{3}{2}}}
\int_{0}^{+\infty}d|\delta ||\delta |^2e^{-\frac{3|\delta |^2}{2W^2}}
\frac{G_2^{-1}}{G_1^{-1}G_2^{-1}-|\delta |^2}
\label{mG}
\end{equation}
This integral form can be easily represented \cite{Sch} by continuous
fraction (\ref{Gk}), (\ref{Sig_k}) with $\kappa =0$ with combinatorial
coefficients $s(k)$, defined in (\ref{sHeis}).

\begin{figure}
\epsfxsize=14cm
\epsfysize=18cm
\epsfbox{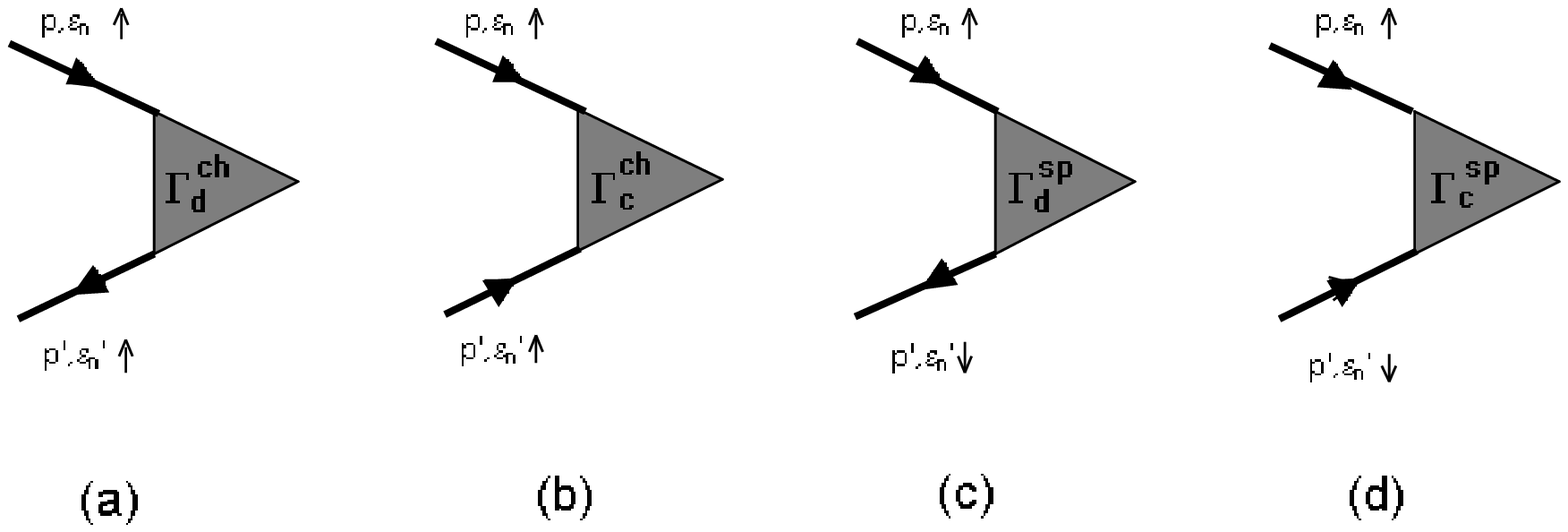}
\caption{Vertex parts with different diagram combinatorics.}
\label{vertex}
\end{figure}

Situation with combinatorial factors $r(k)$ for two -- particle vertices
is more complicated. In fact we have to consider four types of vertices,
shown in Fig. \ref{vertex}. For all these vertices the recursion procedure
has the form of (\ref{Gamma}), but signs and combinatorial factors $r(k)$ are
different. Consider first all these vertices for an electron in coherent
field ${\bf\delta}$.

1. Charged vertex (spin projection is conserved) in diffusion (particle -- hole)
channel (Fig. \ref{vertex} (a)):
\begin{equation}
\Gamma ^{ch}_d=\sum_{i,\sigma}G_{1\uparrow ;i\sigma}G_{i'\sigma ;1'\uparrow}=
\frac{(G_2G_{2'})^{-1}+|\delta |^2}{d_\delta }
\label{Gam_chd}
\end{equation}
where $i$ and $\sigma$ take the values 1,\ 2 and $\uparrow , \downarrow$, 
and we introduced notation $(\varepsilon '_n, {\bf p}')\to 1'$,
$(\varepsilon '_n, {\bf p}'+{\bf Q})\to 2'$ ¨
$d_\delta =[(G_1G_2)^{-1}-|\delta |^2][(G_{1'}G_{2'})^{-1}-|\delta |^2]$

2. Charged vertex in Cooper (particle -- particle) channel\footnote{
It appears in case triplet pairing.} (Fig. \ref{vertex} (b)):
\begin{equation}
\Gamma ^{ch}_c=\sum_{i,\sigma}G_{1\uparrow ;i\sigma}G_{1'\uparrow ;i'\sigma}=
\frac{(G_2G_{2'})^{-1}+\delta _l^2}{d_\delta }
\label{Gam_chc}
\end{equation}

3. Spin vertex (spin projection changes sign) in diffusion (particle -- hole)
channel (Fig. \ref{vertex} (c)):
\begin{equation}
\Gamma ^{sp}_d=\sum_{i,\sigma}G_{1\uparrow ;i\sigma}G_{i'-\sigma ;1'\downarrow}=
\frac{(G_2G_{2'})^{-1}-\delta _l^2}{d_\delta }
\label{Gam_spd}
\end{equation}

4. Spin vertex in Cooper (particle -- particle) channel (Fig. \ref{vertex} (d)):
\begin{equation}
\Gamma ^{sp}_c=\sum_{i,\sigma}G_{1\uparrow ;i\sigma}G_{1'\downarrow ;i'-\sigma}=
\frac{(G_2G_{2'})^{-1}+(|\delta _t|^2-\delta _l^2)}{d_\delta }
\label{Gam_spc}
\end{equation}
Physical vertices are obtained from these by averaging by fluctuations of
coherent field ${\bf\delta }$ with the help of (\ref{SpFl2}).

Finally, we can see that the vertex $\Gamma ^{ch}_d$ is determined by 
Eq. (\ref{Gam_chd}), while all other vertices take the form\footnote
{This form is equivalent to (\ref{Gam_chc}),(\ref{Gam_spd}),(\ref{Gam_spc}) 
after averaging}:
\begin{equation}
\Gamma =
\frac{(G_2G_{2'})^{-1}\pm\frac{1}{3}|\delta |^2}{d_\delta }
\label{GamN}
\end{equation}
where ``$+$'' corresponds to vertices $\Gamma ^{ch}_c$ and $\Gamma ^{sp}_c$,
while ``$-$'' is taken for $\Gamma ^{sp}_d$.

In case of the vertex $\Gamma ^{ch}_d$ we have, obviously, $r(k)=s(k)$. 
It follows directly from the fact that diagram expansion for the physical vertex
$<\Gamma ^{ch}_d>$ can be obtained by inserting of appropriate bare vertex
into all electron lines in any arbitrary diagram for the single -- particle
Green's function. This insertion does not change the direction of either 
electronic line or spin projection, so diagram combinatorics does not change.

In the limit of infinite correlation length any ``skeleton'' diagram for
the vertex differs from ``ladder'' diagram of the same order in interaction
$\frac{W^2}{3}\delta ({\bf q}-{\bf Q})$ only by sign and the factor of
$2^p$, where $p$ --- is the number of wavy interaction lines. Thus, the sum of
all ``skeleton'' diagrams in the given order reduces to the contribution of the
appropriate ``ladder'' diagram with interaction 
$\frac{W^2}{3}\delta ({\bf q}-{\bf Q})$, multiplied by combinatorial factor,
which we call the ``number'' of ``skeleton'' diagrams of the given order.

First term in Eqs. (\ref{Gam_chd})-(\ref{Gam_spc}) is the same for all vertices
and generates, after averaging procedure, the ``numbers'' of ``skeleton''
diagrams of even order in $W^2$ (this term corresponds to $i=1$ contribution
in these expressions). We see that the ``numbers'' of ``skeleton'' diagrams
of even order are the same for all types of vertices.
The second term in these expressions generates ``numbers'' of diagrams of
odd order (corresponding to contributions with $i=2$). Accordingly, the
``numbers'' of ``skeleton'' diagrams of odd order for three vertices,
defined by (\ref{GamN}), is equal to $\pm \frac{1}{3}$ of appropriate 
``numbers'' for the vertex $\Gamma ^{ch}_d$. The sign ``$-$'', corresponding
to the vertex $\Gamma ^{sp}_d$, may be compensated by appropriate change of
sign in recursion procedure for this vertex. It follows that the sign before
the second term in (\ref{GamN}) determines the sign in recursion procedure
(\ref{Gamma}) for these vertices, while combinatorial factors $r(k)$ for all
three vertices are the same.

The ``number'' of ``skeleton'' diagrams of the order $L$ is now\footnote{
The factor of $3^L$ appears due to the fact that both recursion procedure
(\ref{Gamma}) and combinatorial coefficients $r(k)$ correspond to expansion
in powers of $W^2$, while the ``number'' of ``skeleton'' diagrams is
defined for an expansion in powers of $\frac{W^2}{3}$.}:
\begin{equation}
3^L\prod_{1\leq k\leq L}r(k)
\label{SkNum}
\end{equation}
Thus, for even $L=2n$ we obtain:
\begin{equation}
\prod_{1\leq k\leq 2n}r(k)=\prod_{1\leq k\leq 2n}s(k)
\label{Num_ev}
\end{equation}
For odd $L=2n+1$:
\begin{equation}
\prod_{1\leq k\leq 2n+1}r(k)=\frac{1}{3}\prod_{1\leq k\leq 2n}s(k)
\label{Num_od}
\end{equation}
Then, taking into account (\ref{sHeis}), we immediately obtain (\ref{rk}).

In this paper we were mainly interested in the vertex $\Gamma ^{sp}_c$.
The analysis, performed above, shows that for this vertex we get the
recursion procedure without alternating signs for the case of 
$s$-wave pairing, when symmetry factor $e({\bf p})$ of the vertex is equal
to $1$. In case of $d$-wave pairing, when superconducting gap changes sign
after change of the momentum by ${\bf Q}$ (i.e. $e({\bf p})=
-e({\bf p}+{\bf Q}))$), the sign in recursion procedure is changed to the
opposite \cite{KSS} and we obtain recursion procedure with alternating signs.
Note, that in the case of Ising -- like spin fluctuations, analyzed in
Ref. \cite{KSS}, situation with signs in recursion procedure for the vertex is 
strictly opposite. This fact is easily understood from an expression
(\ref{Gam_spc}) for the vertex $\Gamma ^{sp}_c$. In Ising case two transverse
components (i.e. the field $\delta _t$) just disappear, leading to the sign
change of the second term in (\ref{Gam_spc}) and in recursion procedure.

\newpage

\newpage

\newpage

\newpage

\newpage

\newpage

\newpage

\newpage

\newpage

\newpage

\newpage

\newpage

\newpage

\newpage

%\newpage

\newpage

\newpage


\begin{thebibliography}{99}

%\begin{references}

\bibitem{Lor}J.L.Tallon, J.W.Loram. Physica C {\bf 349}, 53 (2000)
\bibitem{MS}M.V.Sadovskii.\ Physics Uspekhi {\bf 44}, 515 (2001)
\bibitem{Pines}D.Pines.\ ArXiv:\ cond-mat/0404151
\bibitem{Sch} J.Schmalian, D.Pines, B.Stojkovic.\  Phys.Rev. {\bf B60}, 667 
(1999)
\bibitem{KS}E.Z.Kuchinskii, M.V.Sadovskii. JETP {\bf 88}, 968 (1999)
\bibitem{KSS}E.Z.Kuchinskii, M.V.Sadovskii, N.A.Strigina. JETP {\bf 98}, 748
(2004) 
\bibitem{KK}N.A.Kuleeva, E.Z.Kuchinskii. Fiz. Tverd. Tela {\bf 46}, 1557 (2004) 
\bibitem{SS}M.V.Sadovskii, N.A.Strigina. JETP {\bf 95}, 526 (2002)
\bibitem{AGD}A.A.Abrikosov, L.P.Gorkov, I.E.Dzyaloshinskii.
Methods of Quantum Field Theory in Statistical Physics. Pergamon Press, 
Oxford, 1965
\bibitem{PS}A.I.Posazhennikova, M.V.Sadovskii. JETP Lett. {\bf 63}, 358
(1996);\ JETP {\bf 85}, 1162 (1997)
\bibitem{Radt}R.J.Radtke, K.Levin, H.-B.Sch\"uttler, M.R.Norman.
Phys.Rev. B {\bf 48}, 653 (1993)
\bibitem{AC}A.Posazhennikova, P.Coleman. Phys.Rev. {\bf B67}, 165109 (2003)
\bibitem{NT}S.H.Naqib, J.R.Cooper, J.L.Tallon, R.S.Islam, R.A.Chakalov.
ArXiv:cond-mat/0312443
\bibitem{PT}M.R.Presland, J.L.Tallon, R.G.Buckley, R.S.Liu, N.E.Flower.\ 
Physica C {\bf 176}, 95 (1991)
\bibitem{Gosch}A.E.Karkin, S.A.Davydov, B.N.Goshchitskii, S.V.Moshkin, 
M.Yu.Vlasov.\ Fiz. Metals -- Metallogr. {\bf 76}, 103 (1993)
\bibitem{Uch}Y.Fukuzumi, K.Mizuhashi, K.Takenaka, S.Uchida.
Phys.Rev.Lett. {\bf 76}, 684 (1996)
\bibitem{Tall}J.L.Tallon, C.Bernhard, G.V.M.Williams, J.W.Loram.
Phys.Rev.Lett. {\bf 79}, 5294 (1997)
\bibitem{Tolp}S.K.Tolpygo, J.-Y.Lin, M.Gurvitch, S.Y.Hou, J.M.Phillips.
Phys.Rev. B {\bf 53}, 12454, 12462 (1996)
\bibitem{Rull}F.Rullier-Albenque, H.Alloul, R.Tourbot. Phys.Rev.Lett. {\bf 91}, 
047001 (2003)
\bibitem{Scloc}M.V.Sadovskii. Superconductivity and Localization.
World Scientific, Singapore 2000
\bibitem{PosSad}A.I.Posazhennikova, M.V.Sadovskii. JETP Lett. {\bf 65},
270 (1997)
\bibitem{HN}G.Haran, A.D.S.Nagy. Phys.Rev. {\bf 54}, 15463 (1996)
\bibitem{Valla}T.Valla, A.V.Fedorov, P.D.Johnson, Q.Li, G.D.Gu, N.Koshizuka.
Phys.Rev.Lett. {\bf 85}, 828 (2000)
\bibitem{Kam}A.Kaminski, H.M.Fretwell, M.R.Norman, M.Randeria, S.Rosenkranz,
J.C.Campuzano, J.Mesot, T.Sato, T.Takahashi, T.Terashima, M.Takano,
K.Kadowaki, Z.Z.Li, H.Raffy. Arxiv: cond-mat/0404385
\bibitem{KS01}E.Z.Kuchinskii, M.V.Sadovskii. JETP {\bf 90}, 535 (2000)
\bibitem{KS02}E.Z.Kuchinskii, M.V.Sadovskii. JETP {\bf 94}, 654 (2002)
\bibitem{Pan}S.H.Pan, J.P.O'Neil, R.L.Badzey, C.Chamon, H.Ding,
J.R.Engelbrecht, Z.Wang, H.Eisaki, S.Uchida, A.K.Gupta. Nature {\bf 413},
282 (2001)
\bibitem{McE}K.McElroy, D.-H.Lee, J.E.Hoffman, K.M.Lang, E.W.Hudson, H.Eisaki,
S.Uchida, J.Lee, J.C.Davis. ArXiv: cond-mat/0404005
\bibitem{Kap}A.Fang, C.Howald, N.Kanenko, M.Greven, A.Kapitulnik.
ArXiv: cond-mat/0404452


%\end{references}

\end{thebibliography}
\end{document}